\documentclass[aps,prb,secnumarabic,amsmath,amssymb,reprint]{revtex4-2}

\usepackage[caption=false]{subfig}
\usepackage[title]{appendix}
\usepackage{tikz}
\usepackage{graphicx}
\usepackage{dcolumn}
\usepackage{bm}
\usepackage{float}
\usepackage{braket}
\usepackage{bbold}
\usepackage{enumerate}
\usepackage{wrapfig}
\usepackage{amsmath} 
\usepackage{amssymb} 
\usepackage{seqsplit}
\usepackage{xcolor}
\usepackage{braket}
\usepackage[export]{adjustbox}
\usepackage[shortlabels]{enumitem}
\usepackage{setspace}
\usepackage{tensor}
\usepackage{hyperref}
\def\theYear{\the\year}
\hypersetup{pdfpagemode=FullScreen, colorlinks=true, allcolors=blue}
\DeclareUnicodeCharacter{2009}{\,}

\newcommand{\beq}{\begin{equation}}
\newcommand{\eeq}{\end{equation}}
\newcommand{\bqa}{\begin{eqnarray}}
\newcommand{\eqa}{\end{eqnarray}}
\newcommand{\nn}{\nonumber}

\newcommand{\id}{ \mathbb 1}

\newcommand{\om}{\omega}
\newcommand{\Om}{\Omega}

\newcommand{\Del}{\Delta}

\newcommand{\gam}{\gamma}
\newcommand{\Gam}{\Gamma}

\newcommand{\sig}{\sigma}

\newcommand{\bigp}[1]{\big({#1}\big)}
\newcommand{\Bigp}[1]{\Big({#1}\Big)}
\newcommand{\Biggp}[1]{\Bigg({#1}\Bigg)}

\newcommand{\bigcb}[1]{\big\{{#1}\big\}}

\newcommand{\br}[1]{\{{#1}\}}

\newcommand{\BQIC}{Berkeley Center for Quantum Information and Computation, Berkeley, California 94720 USA}
\newcommand{\DeptPhys}{Department of Physics, University of California, Berkeley, California 94720 USA}
\newcommand{\DeptChem}{Department of Chemistry, University of California, Berkeley, California 94720 USA}
\newcommand{\USCPhys}{Department of Physics and Astronomy, University of Southern California, Los Angeles, California 90089}
\newcommand{\FZJ}{Peter Grünberg Institute of Quantum Control (PGI-8), Forschungszentrum Jülich GmbH, Wilhelm-Johnen-Straße, 52428 Jülich, Germany}
\newcommand{\USCEE}{Department of Electrical and Computer Engineering, University of Southern California, Los Angeles, California 90089}
\newcommand{\Chapman}{Institute for Quantum Studies, Chapman University, Orange, California 92866}
\newcommand{\ChapmanSchmid}{Schmid College of Science and Technology, Chapman University, Orange, California 92866}
\newcommand{\CQIST}{Center for Quantum Information Science and Technology, University of Southern California, Los Angeles, California 90089}

\date{\today}

\begin{document}
	
\title{Stabilizing two-qubit entanglement with dynamically decoupled active feedback}

\author {Sacha Greenfield$^{1,2,3}$}
\author {Leigh Martin$^{4,5}$}
\author{Felix Motzoi$^{4,6,7}$}
\author{K. Birgitta Whaley$^{4,6}$}
\author {Justin Dressel$^{3,8}$}
\author {Eli M. Levenson-Falk$^{1,2,9}$}
\email {elevenso@usc.edu}

\affiliation{$^1$\USCPhys}
\affiliation{$^2$\CQIST}
\affiliation{$^3$\Chapman}
\affiliation{$^4$\BQIC}
\affiliation{$^5$\DeptPhys}
\affiliation{$^6$\DeptChem}
\affiliation{$^7$\FZJ}
\affiliation{$^8$\ChapmanSchmid}
\affiliation{$^9$\USCEE}
\date{\today}

\begin{abstract}
We propose and analyze a protocol for stabilizing a maximally entangled state of two noninteracting qubits using active state-dependent feedback from a continuous two-qubit half-parity measurement in coordination with a concurrent, non-commuting dynamical decoupling drive. We demonstrate that such a drive can be simultaneous with the measurement and feedback, while also playing a key part in the feedback protocol itself. We show that robust stabilization with near-unit fidelity can be achieved even in the presence of realistic nonidealities, such as time delay in the feedback loop, imperfect state-tracking, inefficient measurements, dephasing from $1/f$-distributed qubit-frequency noise, and relaxation. We mitigate feedback-delay error by introducing a forward-state-estimation strategy in the feedback controller that tracks the effects of control signals already in transit. More generally, the steady state is globally attractive without the need for ancillas, regardless of the error state, in contrast to most known feedback and error correction schemes. 
\end{abstract}

\maketitle

\section{Introduction}
Maximally entangled multi-qubit states are important resources for quantum information processing tasks. However, entangled states decohere more rapidly than single-qubit states, and so must either be generated on demand when needed or stored using some form of stabilization that prevents degradation. Circuit-based quantum error correction protocols are examples of such active entanglement stabilization \cite{lidar2013quantum}, but are resource-intensive since they require additional ancillary qubits, periodic entangling unitary gates, and single-qubit measurements. Bath engineering protocols can stabilize entanglement \cite{shankar2013autonomously,kimchi2016stabilizing,tacchinoSteadyStateEntanglement2018,maStabilizingBellStates2019a,maCouplingmodulationMediatedGeneration2021,santosGenerationMaximallyEntangled2023}, but typically require custom device designs to select the state. 

A natural non-unitary tool to consider for such an entanglement stabilization task is the direct measurement of a two-qubit observable, such as parity \cite{ruskovEntanglementSolidstateQubits2003,williamsEntanglementGenesisContinuous2008}. Periodic or continuous measurements of such an observable can prepare \cite{risteDeterministicEntanglementSuperconducting2013a} and then stabilize \cite{kerckhoffDesigningQuantumMemories2010,mohseniniaAlwaysOnQuantumError2020a,livingstonExperimentalDemonstrationContinuous2022a} the entangled eigenspaces of the measurement and can perform multi-qubit operations \cite{hacohen-gourgyIncoherentQubitControl2018b,blumenthalDemonstrationUniversalControl2022}, even in cases where the qubits do not directly interact \cite{rochObservationMeasurementinducedEntanglement2014,motzoiremotetheory2015, dickel2018chip, kerckhoffremotetheory2009, chantasriQuantumTrajectoriesTheir2016,hacohen-gourgyContinuousMeasurementsControl2020,martinSingleshotDeterministicEntanglement2019,lewalleEntanglementpreservingLimitCycles2020}. Stabilizing a specific state, however, generally requires bath engineering or additional feedback control \cite{wiseman2010quantum,schirmerStabilizingOpenQuantum2010,brif2010control,zhang2017quantum,delangeReversingQuantumTrajectories2014,sayrinRealtimeQuantumFeedback2011a} to break the symmetry of the measurement toward the desired outcome. 

Implementing robust entanglement stabilization using measurement feedback is challenging in practice. The stabilized state should not only be a steady state under ideal conditions for the measurement and other engineered decoupling drives \cite{yamamoto2007feedback,mirrahimiStabilizingFeedbackControls2007,yamamoto2008avoiding,martinDeterministicGenerationRemote2015,govia2022stabilizing,brown2022trade,doucet2023scalable}, but remain stabilized in the presence of realistic nonidealities \cite{motzoiBackactiondrivenRobustSteadystate2016}. Some dynamical states such as single-qubit Rabi oscillations \cite{vijayStabilizingRabiOscillations2012} are less sensitive to environmental noise and so can be more easily stabilized, but entangled states are more susceptible to decoherence, which can drastically lower the achievable steady-state entanglement \cite{rochObservationMeasurementinducedEntanglement2014, motzoiremotetheory2015, dickel2018chip}. Dynamical decoupling can reduce dephasing and prolong coherence \cite{pokharel2018demonstration,tripathi2022suppression,pokharel2023demonstration}, but does not produce true stabilization \cite{khodjasteh2011limits} and is typically separated from the measurement and feedback. Time delays in the feedback loop \cite{kashima2009control,liu2021two} and measurement inefficiency are also unavoidable in the laboratory \cite{amini2013feedback}, and can be disastrous to the stabilization fidelity if not accounted for in the protocol \cite{ahnContinuousQuantumError2002,martinDeterministicGenerationRemote2015,pattiLinearFeedbackStabilization2017}.

In this work, we show that it is possible to combine measurement, a feedback drive, and a dynamical decoupling drive simultaneously to achieve robust entanglement generation and stabilization. As a case study we revisit the simple measurement-feedback protocol introduced in Ref.~\cite{martinDeterministicGenerationRemote2015}, which stabilizes a maximally entangled two-qubit Bell state with feedback from a half-parity measurement. This protocol was vulnerable to dephasing which would mix the target state with a degenerate measurement eigenstate, thus eliminating stability. We expand the protocol to include a concurrent drive that provides both dynamical decoupling \cite{lidar2014review,roy2011storing,franco2014preserving,mccourt2023learning} and acts to remove an undesired fixed point of the dynamics. We analyze both the optimum settings and resulting performance in the presence of experimental nonidealities including the time delay of the feedback loop, measurement inefficiency, and environmentally-induced relaxation and dephasing. Notably, we model environmental dephasing as more realistic non-Markovian qubit-frequency noise from environmental fluctuators, with either white or $1/f$ noise spectra \cite{paladinoNoiseImplicationsSolidstate2014}, which we compare to the standard (Lindblad) Markovian techniques for modeling dephasing. 

Our modified stabilization protocol benefits from two key innovations. First, our dynamical decoupling drive notably differs from standard approaches since it neither acts only during idle times nor acts only on auxiliary subspaces. Instead, the decoupling drive is always on as a second drive that can optionally be adjusted by the feedback controller. This drive both reduces dephasing and provides an active pathway for stabilization out of erroneous states. Second, we develop a forward-state-estimation procedure for the feedback controller that accounts for the time delay in the feedback loop by anticipating the evolution occurring from signals already in transit. We find through numerical simulations that the inclusion of these features significantly improves the performance of the stabilization protocol. For dephasing from $1/f$ fluctuator noise, in particular, the modified protocol can robustly achieve near-unit asymptotic fidelity to the target state in a wide range of conditions. 


\section{State Stabilization Protocol}
\label{sec:protocol}
As in Refs.~\cite{rochObservationMeasurementinducedEntanglement2014,martinDeterministicGenerationRemote2015}, our system is comprised of two qubits that are not interacting, described by a joint (system) density operator $\rho_\text{sys}(t)$. Each qubit experiences environmental dephasing at an average rate $\Gamma_2$, detailed later. We assume a rotating frame, independent control drives, and coupling to a common half-parity measuring device (e.g., \cite{risteDeterministicEntanglementSuperconducting2013a,rochObservationMeasurementinducedEntanglement2014,livingstonExperimentalDemonstrationContinuous2022a}).

The half-parity observable $\hat N = (\hat \sig_z^{(1)} + \hat \sig_z^{(2)})/2$ has three eigenvalues: -1, 0, and 1. The eigenvalues -1 and 1 have the even-parity eigenvectors $\ket{11}$ and $\ket{00}$, respectively, while the eigenvalue 0 has a degenerate odd-parity subspace spanned by $\ket{01}$ and $\ket{10}$. The two odd-parity entangled Bell states, $\ket{\psi_\pm} \equiv \bigp{\ket{01} \pm \ket{10}} /\sqrt{2}$ are thus preserved by the measurement, while the even-parity Bell states $\ket{\phi_\pm} \equiv \bigp{\ket{00} \pm \ket{11}}/\sqrt2$ get disturbed.

\emph{The goal of the protocol is to use the two single-qubit control drives and the joint half-parity measurement to stabilize one maximally entangled Bell state: $\ket{\psi_+}$.}

To use feedback to accomplish this goal, we apply control drives that depend upon the collected measurement record $r(t)$, which has the approximate form,
\begin{equation}
	r(t) dt =\text{Tr}(\rho_\text{sys}(t) \hat N) dt + \sqrt{\tau} ~dW,
	\label{eq:ri}
\end{equation}
where $dW^2 = dt$ is a stochastic Wiener increment \cite{wiseman2010quantum} and $\tau = 1/2\eta\Gamma$ is a timescale set by the measurement rate $\eta\Gamma$, given efficiency $\eta\in[0,1]$ and measurement-dephasing rate $\Gamma$. This stochastic process models, e.g., a homodyne measurement of $\hat{N}$ after phase-sensitive amplification of the informational quadrature, then shifting and rescaling to normalize the mean signal to the eigenvalues of $\hat{N}$ \cite{korotkovQuantumBayesianApproach2011}.

We synchronize the applied single-qubit drives to be either corotating or counterrotating, $\hat{\sigma}_y^{(1)}\pm\hat{\sigma}_y^{(2)} \equiv \hat{Y}_\pm$, yielding effective two-qubit drive Hamiltonians,
	\begin{align}\label{eq:H-omega-delta}
		\hat H_\Om(t) &= \hbar\Om ~ \hat Y_{+}/2, &
		\hat H_{\Delta}(t) &= \hbar\Del ~ \hat Y_{-}/2, 
	\end{align}
with amplitudes $\Omega$ and $\Delta$ that are generally functions of time $t$, the collected record $r(t-\tau_d)$ after a delay $\tau_d$ from the feedback loop, and/or an estimated state $\rho_\text{est}(t)$. Importantly, both $\hat{N}$ and $\hat H_\Del$ act trivially on the stabilized $\ket{\psi_+}$, meeting the condition for global convergence \cite{schirmerStabilizingOpenQuantum2010, motzoiBackactiondrivenRobustSteadystate2016}. The corotating drive $\hat H_\Om$ with matched drive phases generates rotations in the $\bigcb{\ket{\psi_+}, \ket{\phi_-}}$ subspace that contains the target state $\ket{\psi_+}$, making it the \emph{feedback drive} \cite{martinDeterministicGenerationRemote2015}. The counterrotating drive $\hat H_\Del$ with relative $\pi$-phase shift generates rotations in the $\bigcb{\ket{\phi_+}, \ket{\psi_-}}$ subspace orthogonal to $\ket{\psi_+}$; these rotations also suppress slowly-varying dephasing noise on $\ket{\psi_+}$, making this the \emph{dynamical decoupling drive}. 

\begin{figure}
	\centering
	\includegraphics[width=0.96\linewidth]{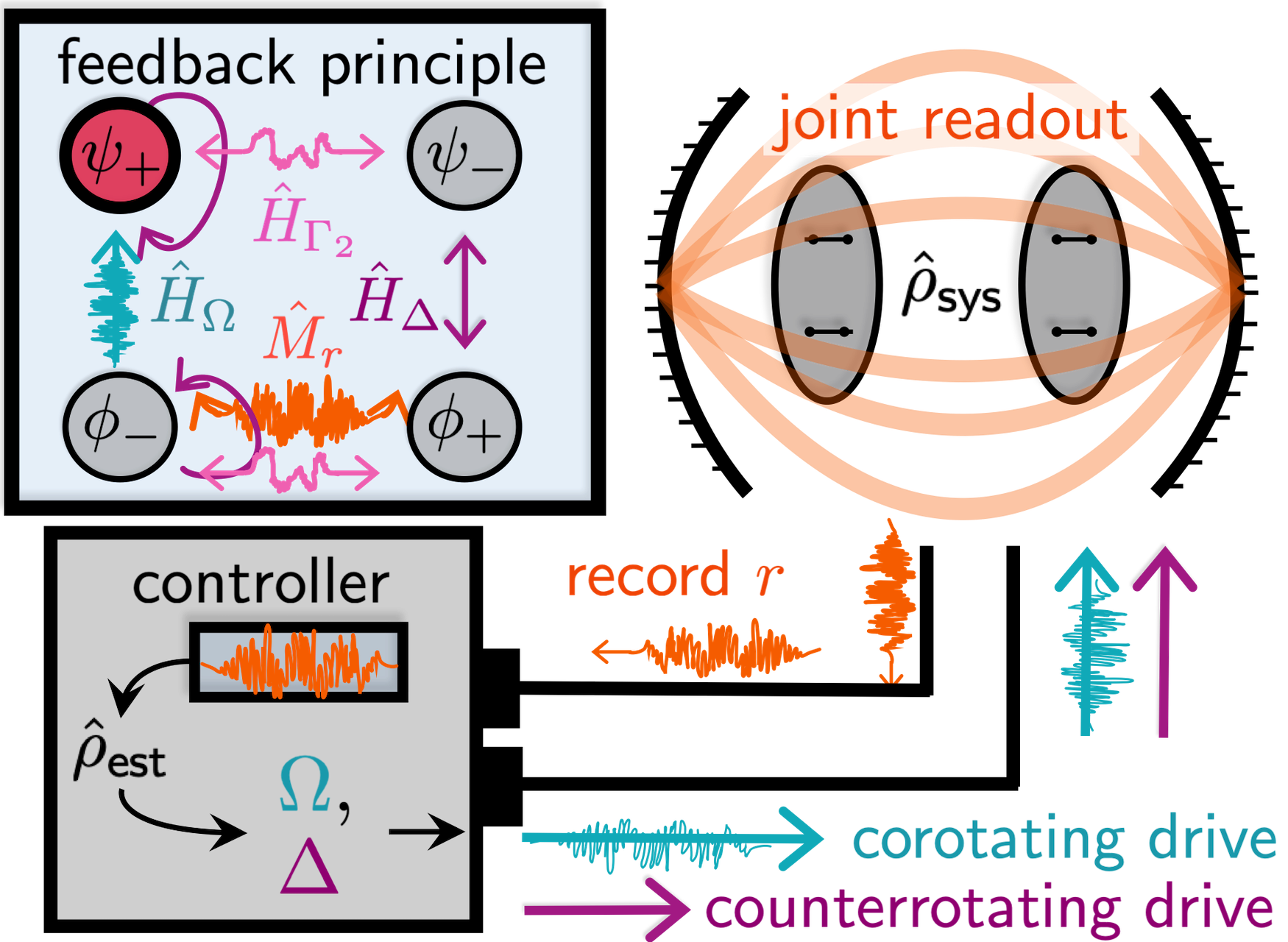}
	\caption{Schematic for stabilizing the entangled Bell state $\ket{\psi_+}$ (red). Two noninteracting qubits $\rho_\text{sys}(t)$ undergo a half-parity measurement with dephasing rate $\Gamma$ (orange), yielding a noisy record $r$ and stochastic transfer $\hat{M}_r$ between $\ket{\phi_+}\leftrightarrow\ket{\phi_-}$ while stabilizing $\ket{\psi_\pm}$. Environmental dephasing at rate $\Gam_2$ (pink) also causes stochastic transfer $\hat{H}_{\Gamma_2}$ between $\ket{\phi_+}\leftrightarrow\ket{\phi_-}$ and $\ket{\psi_+}\leftrightarrow\ket{\psi_-}$. A controller uses the measurement record to internally track an estimated state $\rho_\text{est}(t)$, then applies a feedback drive $\hat H_\Om$  (blue) after a delay $\tau_d$ at rate $\Omega[\rho_\text{est}(t)]$ that induces a biased transfer $\ket{\phi_-}\rightarrow \ket{\psi_+}$, as well as a dynamical decoupling drive $\hat H_\Del$ (purple) at rate $\Delta[\rho_\text{est}(t)]$ that stabilizes $\ket{\psi_+}$ and $\ket{\phi_-}$ and also transfers $\ket{\psi_-}\leftrightarrow\ket{\phi_+}$.}
	\label{fig:state-diagram}
\end{figure}

Fig.~\ref{fig:state-diagram} shows a conceptual schematic of how the two drives and the measurement both prepare and stabilize the entangled state $\ket{\psi_+}$. After preparing the product state $|+,+\rangle = \bigp{\ket{\psi_+} + \ket{\phi_+}}/\sqrt2$, the measurement stabilizes $\ket{\psi_+}$ at dephasing-rate $\Gamma$ but stochastically transfers within the subspace $\ket{\phi_+} \leftrightarrow \ket{\phi_-}$. The feedback drive unitarily rotates between $\ket{\phi_-} \leftrightarrow \ket{\psi_+}$ at rate $\Omega$, but the measurement feedback law biases this rotation to accumulate population in $\ket{\psi_+}$. Meanwhile, dephasing induces transfers between $\ket{\psi_+} \leftrightarrow \ket{\psi_-}$ and $\ket{\phi_+} \leftrightarrow \ket{\phi_-}$ at rate $\Gamma_2$, with $\ket{\psi_-}$ invisible to both $\Omega$ and the measurement. Thus, $\ket{\psi_-}$ is an undesired fixed point of the protocol so far. However, the decoupling drive at rate $\Delta$ unitarily rotates between $\ket{\psi_-}\leftrightarrow\ket{\phi_+}$, providing a pathway through $\Gamma$ and $\Omega$ back to $\ket{\psi_+}$. Neither the dephasing $\Gamma_2$ nor the decoupling $\Delta$ were considered in Ref.~\cite{martinDeterministicGenerationRemote2015}, nor in similar ancilla-free protocols in \cite{risteDeterministicEntanglementSuperconducting2013a,martinSingleshotDeterministicEntanglement2019,lewalleEntanglementpreservingLimitCycles2020,mirrahimiStabilizingFeedbackControls2007}.

\subsection{Modeling the Feedback Controller}
\label{subsec:controller}
Simulations of real-time measurement feedback must distinguish between the true (but inaccessible) state of the system $\rho_\text{sys}(t)$ and the estimated state $\rho_\text{est}(t)$ being internally tracked by the controller, both indicated in Fig.~\ref{fig:state-diagram}. Simulations must evolve both $\rho_\text{sys}(t)$ to determine the measurement record in Eq.~\eqref{eq:ri} and $\rho_\text{est}(t)$ to determine the feedback drives in Eqs.~\eqref{eq:H-omega-delta} \cite{pattiLinearFeedbackStabilization2017,szigetiRobustnessSystemfilterSeparation2013}. 

The evolution of the internal state $\rho_\text{est}(t)$ has three key differences. First, the controller does not know the specific environmental fluctuations causing ensemble decoherence, so can only use an average decoherence model. Second, the controller only has the collected record $r(t)$ up to a delay $\tau_d$ before the output feedback signal will reach the system state $\rho_\text{sys}(t+\tau_d)$. 
Third, whereas the system state evolves with simulation timestep $dt$, the controller estimate $\rho_\text{est}(t)$ evolves with timestep $\delta t = n ~dt$, where $n$ is some integer. This timestep is called a ``control cycle'' and represents the time interval in which the controller performs the Bayesian update, estimates the state, and updates the drive tones. Differentiating the two timesteps is important because, whereas $dt$ must be much smaller than any other simulation timescale to maintain accuracy of the discrete state-update, $\delta t$ will be experimentally constrained by FPGA controller latency and may be comparable to other timescales (such as a Rabi period).

\subsection{Evolution With Decoherence and Measurement}
\label{subsec:evolutionDecoherence}
 We model dephasing as stochastic dispersive energy shifts from weakly coupling to a noisy environment \cite{paladinoNoiseImplicationsSolidstate2014}, 
	 \begin{equation}
	\hat H_\omega(t) = \hbar\omega\bigp{\chi_1(t)\, \hat \sigma_z^{(1)} +  \chi_2(t)\, \hat \sigma_z^{(2)}}/2.
	\label{eq:stochastic-ham}
\end{equation}	
Here $\chi_{1,2}(t)$ are independent zero-mean stochastic processes, which we generate with either white or $1/f$ power spectral densities to encompass the two extremes of commonly observed frequency noise. From here on we refer to these as ``white noise'' and ``fluctuator noise''. We calibrate the amplitude $\hbar\omega$ so that ensemble-averaging the fluctuations yields the dephasing rate $\Gam_{fl}$ (Appx. \ref{sec:simulating-dephasing}). 

After measuring $\hat{N}$, the observed record $r(t)$ in Eq.~\eqref{eq:ri} integrated over a small timestep $dt$ is Gaussian-distributed with mean $\text{Tr}(\rho_\text{sys}(t)\hat{N})$ and variance $1/2\eta\Gamma dt$. The induced partial collapse is thus encapsulated by the Kraus operator update \cite{korotkovQuantumBayesianApproach2011,pattiLinearFeedbackStabilization2017},
\begin{align}
	\hat M_{r(t)dt} &= \exp \left[\eta\Gamma\,dt \left(r(t) \hat N - \hat N^2/2\right)\right], 
 \label{eq:measurement-operator}\\
    \mathcal{M}_{r(t)}[\rho(t)] &= \hat{M}_{r(t)}\rho(t)\hat{M}^\dagger_{r(t)} / \text{Tr}\left(\hat{M}^\dagger_{r(t)}\hat{M}_{r(t)}\rho(t)\right).
    \label{eq:bayesian-update}
\end{align}

For inefficient measurements, $\eta<1$, the uncollected part of the signal must be averaged over as unknown, causing residual measurement-dephasing. This dephasing is described with the additional jump/no-jump Lindblad map. We use the same procedure to model relaxation ($T_1$) processes, and group them together into the single map $\mathcal{L}_{dt}[\rho(t)]$ 
\cite{khezriQubitMeasurementError2015,supplement}:
\begin{subequations}
\begin{align}
		\mathcal{L}_{dt}[\rho(t)] &= \hat J_0 \rho(t) \hat J_0^\dagger + \sum_{k=1}^3 (\gam_k  dt ) \hat J_k \rho(t) \hat J_k^\dagger,  
    \\
    \hat J_{1,2} &= \hat \sigma_-^{(1,2)}/\sqrt2, \qquad \qquad \gam_{1,2} = 1/ T_1 
    \\
    \hat J_3 &= \hat N / \sqrt2, \qquad \qquad \gam_3 = (1-\eta)\Gam,
    \\
    \hat J_0 &=  \sqrt{\id -  \sum_{k=1}^3 (\gam_k dt) \hat J_k^\dagger \hat J_k}.
 		\end{align}
 		\label{eq:lindblad-map-main}
  \end{subequations}
 		
The composite update map for the system state is thus,
\begin{align}\label{eq:state-update-main}
    \rho_\text{sys}(t+dt) &= [\mathcal{U}_{dt}\circ\mathcal{L}_{dt}\circ\mathcal{M}_{r(t)dt}][\rho_\text{sys}(t)], 
\end{align}
with the unitary update, $\mathcal{U}_{dt}[\rho(t)] = \hat{U}_{dt,\rm tot}(t)\rho(t)\hat{U}^\dagger_{dt,\rm tot}$,
\begin{align}
    \hat{U}_{dt,\rm tot} &= \exp(dt[\hat{H}_\Delta(t) + \hat{H}_\Omega(t) + \hat{H}_\omega(t)]/i\hbar),
    \label{eq:unitary-total}
\end{align}
recalling that $\Omega$ and $\Delta$ generally depend upon $\rho_\text{est}(t)$.
This completely positive map maintains state integrity on iteration and is accurate to order $dt$, which is chosen to be much smaller than any other period in the system evolution.

The controller update omits the noisy $\hat{H}_\omega$ as unknown and either replaces it with an average Lindblad dephasing map (Appx. \ref{sec:simulating-dephasing}\ref{subsec:effective-dephasing}) or omits it as negligible over $\delta t$. It likewise uses a Lindblad map to model the effects of imperfect measurement efficiency and relaxation. It also performs a Bayesian update based on the measurement record averaged over the control cycle duration $\delta t$, i.e. $\tilde r(t) \equiv (1/n) \sum_{m = 0}^{n-1} r(t - m dt)$, where the full update is given by Eqs. \ref{eq:state-update-main}-\ref{eq:unitary-total} with $\rho_{\rm sys}(t + dt) \rightarrow \rho_{\rm est}(t + \delta t)$, $r(t) \rightarrow \tilde r(t)$, and $dt \rightarrow \delta t$.

\subsection{Forward Estimation}
\label{subsec:forwardEstimation}
Due to the feedback loop delay, the measurement record does not reflect the effects of drives output over the past $\tau_d$. This would cause inaccurate control if $\tau_d > \delta t$, as the system would over-correct by repeatedly trying to rotate the state before the effects of those rotations could be measured. We therefore implement a novel \emph{foward estimation} protocol to correct for the delay. After the measurement update described above, the controller then makes a separate forward estimation by the delay time $\tau_d$. This forward estimation predicts the effects of the control signals already output over the preceding $\tau_d$, but of course does not include any measurement information that the controller would not have access to. For details of the forward estimation procedure, see Appendix \ref{app:forward-estimation}. This $\rho_\text{est}(t+\tau_d)$ is used to determine the new control signals $\Omega(t+\delta t)$ and $\Delta(t+\delta t)$ to output, which are then added to the buffer of emitted control signals used for forward estimation in future control cycles. We show the effectiveness of this forward-estimation procedure in Sec.~\ref{subsec:forwardTests}.

\subsection{Optimized feedback}
The choice of feedback drives $\Omega(t)$ and $\Delta(t)$ in Eq.~\eqref{eq:H-omega-delta} affect the stabilization effectiveness, so we choose them to extremize the fidelity to $\ket{\psi_+}$,
\begin{align}\label{eq:fidelity}
    \mathcal{F}[\rho_\text{est}(t)] &= \bra{\psi_+}\!\rho_\text{est}(t)\!\ket{\psi_+},
\end{align}
after one time step $\delta t$. Imposing the conditions $\partial\mathcal{F}[\rho_\text{est}(t+\delta t)]/\partial \Delta = 0$ and $\partial\mathcal{F}[\rho_\text{est}(t+\delta t)]/\partial \Omega = 0$ yields 
\begin{align}\label{eq:delta-opt}
    2\Delta_{\rm opt}[\rho]~\delta t &= \tan^{-1}[2\text{Re}(\rho_{\phi_+,\psi_-}) / (\rho_{\phi+} - \rho_{\psi-})] \\
    2\Omega_{\rm opt}[\rho]~\delta t &=\tan^{-1}[2\text{Re}(\rho_{\psi_+,\phi_-}) / (\rho_{\psi+} - \rho_{\phi-})]\label{eq:omega-opt}
\end{align}
using the notation $\rho_{\psi,\phi}=\bra{\psi}\!\rho\!\ket{\phi}$ and $\rho_\psi = \rho_{\psi,\psi}$.
Intuitively, feedback coefficients are weighted \emph{coherences} for the subspaces in which the drives and measurement act. In the limit of measurement strength much larger than other noise scales, this strategy produces a steady state that is self-healing for arbitrary error operators.

We also consider a dynamical decoupling strategy with an always-on constant drive $\Delta_{\rm opt} \to \Delta_{\rm d.d.}$. 
This simplification gives less-than-optimal fidelity in the next time step, but keeps the drive agnostic to state estimation errors while increasing decoupling from dephasing (particularly for fluctuators noise) and still enabling the crucial transfer $\ket{\psi_-}\rightarrow\ket{\phi_+}$. The feedback drive $\hat{Y}_+$, however, disturbs the target state $\ket{\psi_+}$, so the optimal drive $\Omega_{\rm  opt}$ is necessarily state-dependent even in this simplified protocol. As shown below, this tradeoff of local optimality for enhanced decoupling is beneficial in some circumstances.

\begin{figure}
	\centering
	\includegraphics[width=\linewidth]{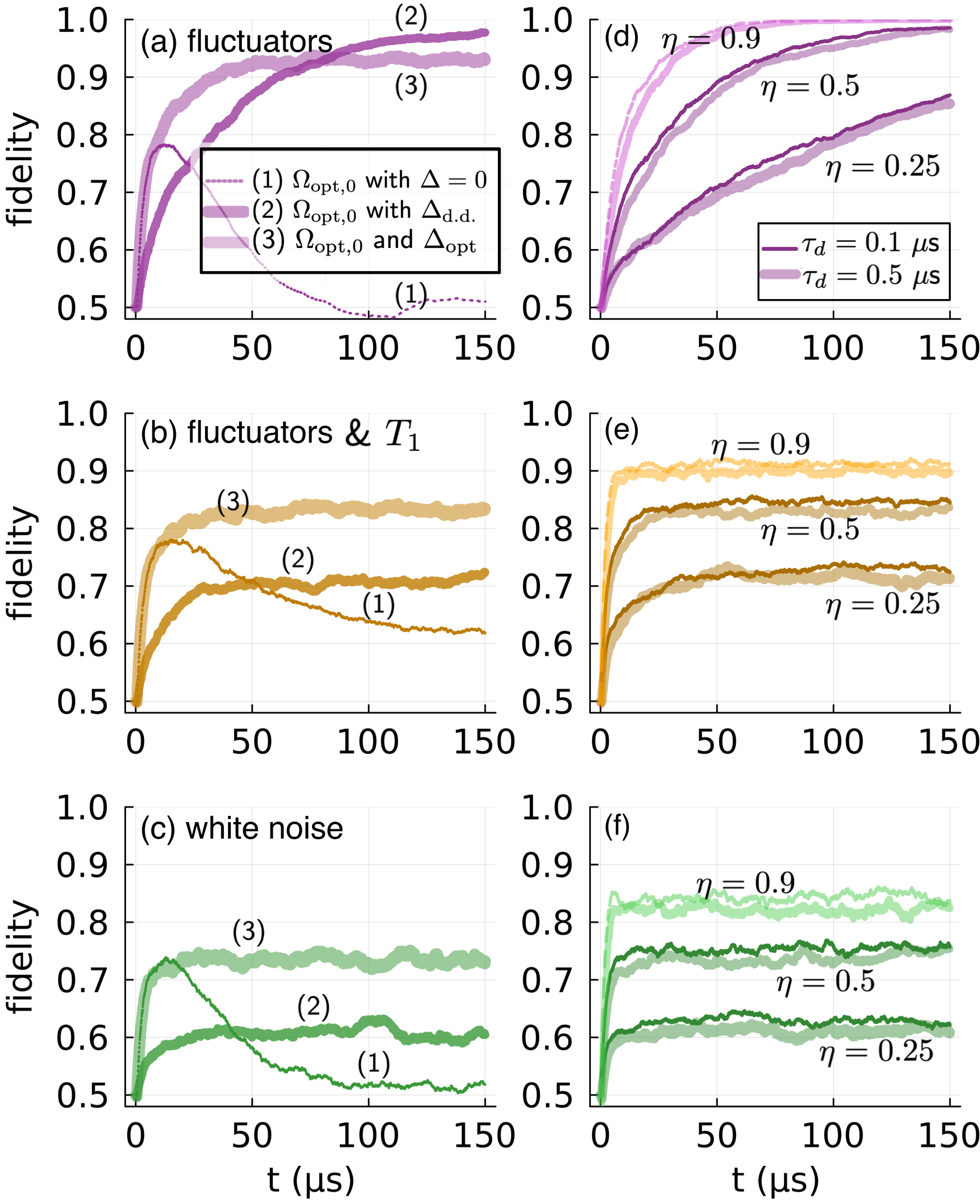}
    \caption{Fidelity  $\mathcal{F}(t)$ of the ensemble-averaged state ($N = 1000$ trajectories) to target $\ket{\psi_+}$, compared across different feedback protocols \textbf{(a-c)} and apparatus parameters \textbf{(d-f)}. The ensemble is initialized in state $\ket{+,+}$ and evolves under optimal corotating drive $\Omega_{\rm opt,0}(\rho)$ and specified counterrotating drive $\Delta$. Except where specified, simulations use measurement-dephasing rate $\Gamma = 1/\mu$s, environmental-dephasing rate $\Gamma_2 = 1/50\mu$s, feedback loop delay $\tau_d = 0.5~\mu$s, and measurement efficiency $\eta = 0.5$.
    \textbf{(a-c)} Comparison of three decoupling strategies, where $\Delta$ is either (1) off, (2) a constant $\Delta_{\rm d.d.}/2\pi = 25$ MHz simultaneous with $\Omega_\text{opt,0}$, or (3) an optimized state-dependent drive $\Delta_\text{opt}(\rho)$ simultaneous with $\Omega_\text{opt,0}$. For dephasing from pure $1/f$ fluctuator noise in \textbf{(a)}, simultaneous feedback and constant counterrotating (decoupling) drive achieves near-unit fidelity asymptotically. For fluctuators and $T_1$ in \textbf{(b)}, the state-dependent counterrotating drive is most effective but saturates to imperfect stabilization. Pure white noise \textbf{(c)} mirrors this result with lower fidelities.  \textbf{(d-f)} Comparison with different measurement efficiencies $\eta$ and feedback loop delays $\tau_d$, using the best control strategy for each noise type.  For $1/f$ noise in \textbf{(d)}, strategy (2) ( constant $\Delta_{\rm d.d.}$) is applied, and nonidealities are observed to slow down stabilization without degrading the asymptotic near-unit fidelity. For fluctuators and $T_1$ in \textbf{(e)} and white noise in \textbf{(f)}, asymptotic fidelities depend on $\eta$ and $\tau_d$. }
    \label{fig:fidelity-timeseries}
\end{figure}

\section{Numerical simulations}
\label{sec:numSims}
We first compare protocols with different combinations of feedback and counterrotating drives across three different noise environments: pure fluctuators noise, fluctuators noise combined with $T_1$ decay, and pure white noise. To compare in a standard way across noise environments, we define the dephasing rate $\Gam_2$ as follows: In the case of pure fluctuators or pure white noise, the total dephasing rate is simply $\Gam_2 = \Gam_{fl}$, i.e. all non-measurement-induced dephasing is due to averaged fluctuations, with $\Gam_{fl}$ calibrated separately (Appx. \ref{sec:simulating-dephasing}\ref{subsec:noise-equivalency}). We assume $T_1 \rightarrow \infty$ for both. In the fluctuators noise combined with $T_1$ decay case, the dephasing is split evenly between the fluctuation-induced and $T_1$-induced dephasing, such that $\Gam_2 = \Gam_{fl} + 1/2T_1$ and $\Gam_{fl} = 1/2T_1 = \Gam_2/2$, with $T_1$ decay described by Eq. \ref{eq:lindblad-map-main}.

\subsection{Optimal Feedback Strategy}
\label{subsec:optimalStrategy}
Fig.~\ref{fig:fidelity-timeseries} shows the time-dependent target-state fidelity from Eq.~\eqref{eq:fidelity} after ensemble-averaging $N = 1000$ simulated trajectories per curve, with time $t\in[0,150]\,\mu$s discretized into bins $dt = 1$~ns, and parameters $\Gamma_2 = 1/50\mu$s, $\Gamma = 1/\mu$s, $\eta = 0.5$, and $\tau_d = 500$ ns. We note that these values are well within the state-of-the-art in the superconducting qubit community, with recently achieved efficiencies of $\eta > 0.7$ \cite{eddinsHighEfficiencyMeasurementArtificial2019,lecocqEfficientQubitMeasurement2021} and dephasing rates of $1/300$ $\mu$s \cite{placeNewMaterialPlatform2021}. Feedback delay time is constrained by the travel time for signals to enter and leave the cryostat and by processing time, leading to typical delay times of $200$-$400$ ns \cite{vijayStabilizingRabiOscillations2012}.  The drives $\Omega$ and  $\Delta$ are updated every $\delta t = 10$ ns control cycle.
In Fig.~\ref{fig:fidelity-timeseries}a-c, we compare three strategies using $\Omega_{\rm opt}$ in Eq.~\eqref{eq:omega-opt}:
\begin{enumerate}
    \item $\Omega_\text{opt}(\rho)$ and $\Delta = 0$, a protocol originally presented in \cite{martinDeterministicGenerationRemote2015};
    \item $\Omega_\text{opt}(\rho)$ and $\Delta_{\rm d.d.}$, where a constant counterrotating drive acts as simultaneous dynamical decoupling and feedback;
    \item $\Omega_\text{opt}(\rho)$ and $\Delta_\text{opt}(\rho)$, where the counterrotating drive choice is locally optimal (Eq.~\ref{eq:delta-opt}).
\end{enumerate}
We comment on these in turn:

\textit{(1) No counterrotating drive.---}
In the presence of environmental noise, the corotating feedback drive $\Omega$ is inadequate for robust stabilization, regardless of noise spectrum ($1/f$ vs. white) or type (pure dephasing or dephasing and $T_1$). Decoherence populates $\ket{\psi_-}$, which is indistinguishable from $\ket{\psi_+}$ under the measurement and inaccessible to the corotating drive. 

\textit{(2) Simultaneous d.d. and feedback.---} For $1/f$ fluctuators noise (Fig. \ref{fig:fidelity-timeseries}a), which cause a non-exponential short-time dephasing (Appx. \ref{sec:simulating-dephasing}), \textit{a state-agnostic constant decoupling drive simultaneous with the corotating feedback is asymptotically the best strategy, achieving near-unit fidelity}. The same approach leads to sub-optimal saturating fidelities for fluctuators and $T_1$ (Fig. \ref{fig:fidelity-timeseries}b) or pure white noise dephasing (Fig. \ref{fig:fidelity-timeseries}c). These results make sense in light of known decoupling properties of $1/f$ noise and white noise \cite{gordonOptimalDynamicalDecoherence2008} (Appx. \ref{sec:simulating-dephasing}). With white noise dephasing or relaxation, the d.d. drive makes the $\ket{\psi_-}$ population accessible to the rest of the feedback loop (enabling some stabilization), but true dynamical decoupling does not occur.

\textit{(3) Locally optimal feedback.---} For noise environments with a white component (Fig. \ref{fig:fidelity-timeseries}b, c), \textit{the locally optimal feedback on both corotating and counterrotating drives is asymptotically the best strategy}. Furthermore, the qualitative similarities between fidelity timeseries for the two noise environments implies that the $T_1$ decay is mainly disruptive via its contribution of $1/2T_1$ to the overall dephasing rate $\Gamma_2$, which makes sense given that decay of individual qubit populations from $\ket1 \mapsto \ket0$ is a non-excitation-conserving map and therefore can be detected by the measurement and subsequently corrected by the feedback. Furthermore, since we model $T_1$ using Lindblad maps, the spectrum of the effective dephasing contribution is white, which further explains the similarities with pure white noise. The protocol performs well, particularly at short times, for pure $1/f$ fluctuators, but is asymptotically sub-optimal if there is no broadband noise.


Fig.~\ref{fig:fidelity-timeseries}d-f show how decreasing measurement efficiency $\eta$ and increasing time delay $\tau_d$ affects stabilization in the three noise environments, using the best feedback strategy for each environment as determined above. Notably, for fluctuators noise, decreasing $\eta$ only slows down the stabilization (due to the resulting decrease in measurement rate $\eta \Gamma$) but does not decrease the asymptotic fidelity. For white noise, as well as for fluctuators noise plus $T_1$ decay, the asymptotic fidelity is a steady state that balances the measurement rate $\eta\Gamma$ and dephasing rate $\Gamma_2$, so it directly depends on $\eta$. These results are in agreement with other work showing d.d.~is most effective against non-white (time-correlated) noise \cite{lidar2014review}.

We note that the locally optimal protocol attempts to optimize fidelity in the next time step, but assumes that the decoherence rate is constant. However, a strong constant d.d. drive greatly reduces the effective dephasing rate in the case of slow ($1/f$) noise. We include this effect in our estimate of the state's dephasing (see Appx.~\ref{subsec:effective-dephasing} for details). When $1/f$ dephasing is dominant, this d.d. approach is globally optimal; when white noise or relaxation are significant, the d.d. is less effective and so the locally optimal protocol is also globally optimal. Future work could explore whether hybrid strategies could be optimal for different ratios of slow and fast decoherence; for instance, a locally optimal protocol with periodic d.d. 

\begin{figure}
	\centering
	\includegraphics[width=0.93\linewidth]{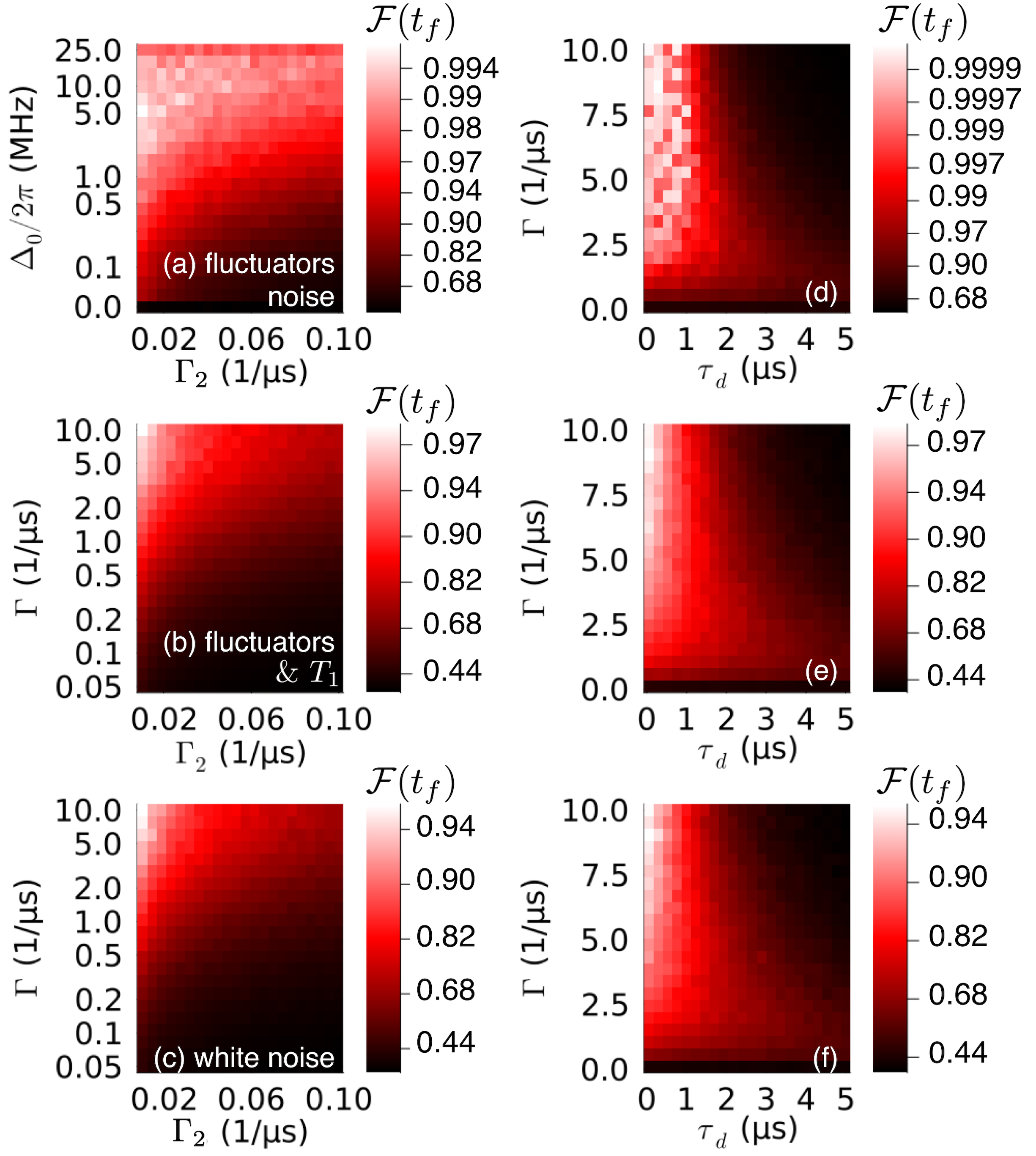}
	\caption{Asymptotic fidelity $\mathcal{F}$ of the ensemble-averaged state ($N = 1000$ trajectories) to target $\ket{\psi_+}$ at $t_f=150\,\mu$s, with color displayed on a log scale. The optimal protocol for each noise type is used. Except where specified as a sweep variable, parameters default to $\eta = 0.5$, $\tau_d = 0.5 ~\mu$s, $\Gamma = 1 /\mu$s, $\Delta_0/2\pi = 25$ MHz, $\Gamma_2 = 1/50.0 ~\mu$s. \textbf{(a)} Fluctuators noise is mitigated with $\Delta_\text{d.d.}$, sweeping the counterrotating drive strength $\Delta_0$, for different dephasing rates $\Gam_2 \in [0.01, 0.1] / \mu$s. Near-unit fidelity $(>0.99)$ is achieved for any dephasing rate with a sufficiently fast decoupling drive $\Delta_0$.
\textbf{(b)}  Fluctuators and $T_1$  and \textbf{(c)}  white noise  are mitigated using $\Delta_\text{opt}$, sweeping the measurement rate $\Gamma$ for different $\Gamma_2$. Fidelities are improved with stronger measurements (larger $\Gamma$), but lag behind the fidelities observed for fluctuators and $\Delta_\text{d.d.}$. \textbf{(d-f)} Measurement rate $\Gamma$ is swept for different feedback loop delay times $\tau_d$ for \textbf{(d)} fluctuators, \textbf{(e)} fluctuators and $T_1$, and \textbf{(f)} white noise. All demonstrate stabilization fidelities that are remarkably robust to loop delay, with short delays showing large optimal $\Gamma$ (up to $10/\mu$s in the range swept) and long delays necessitating smaller $\Gamma$. For the realistically achievable $\Gamma_2 = 1/50.0~\mu$s and $\tau_d = 0.52~\mu$s, optimal fidelities achieved are \textbf{(d)} $\mathcal{F} = 0.9999$ at $\Gam = 5.76/\mu$s, \textbf{(e)}  $\mathcal{F} = 0.95$ at $\Gam = 10.0/\mu$s, and \textbf{(f)}  $\mathcal{F} = 0.92$ at $10.0/\mu$s.}
	\label{fig:heatmaps}
\end{figure}

\begin{figure}
	\centering
	\includegraphics[width=0.8\linewidth]{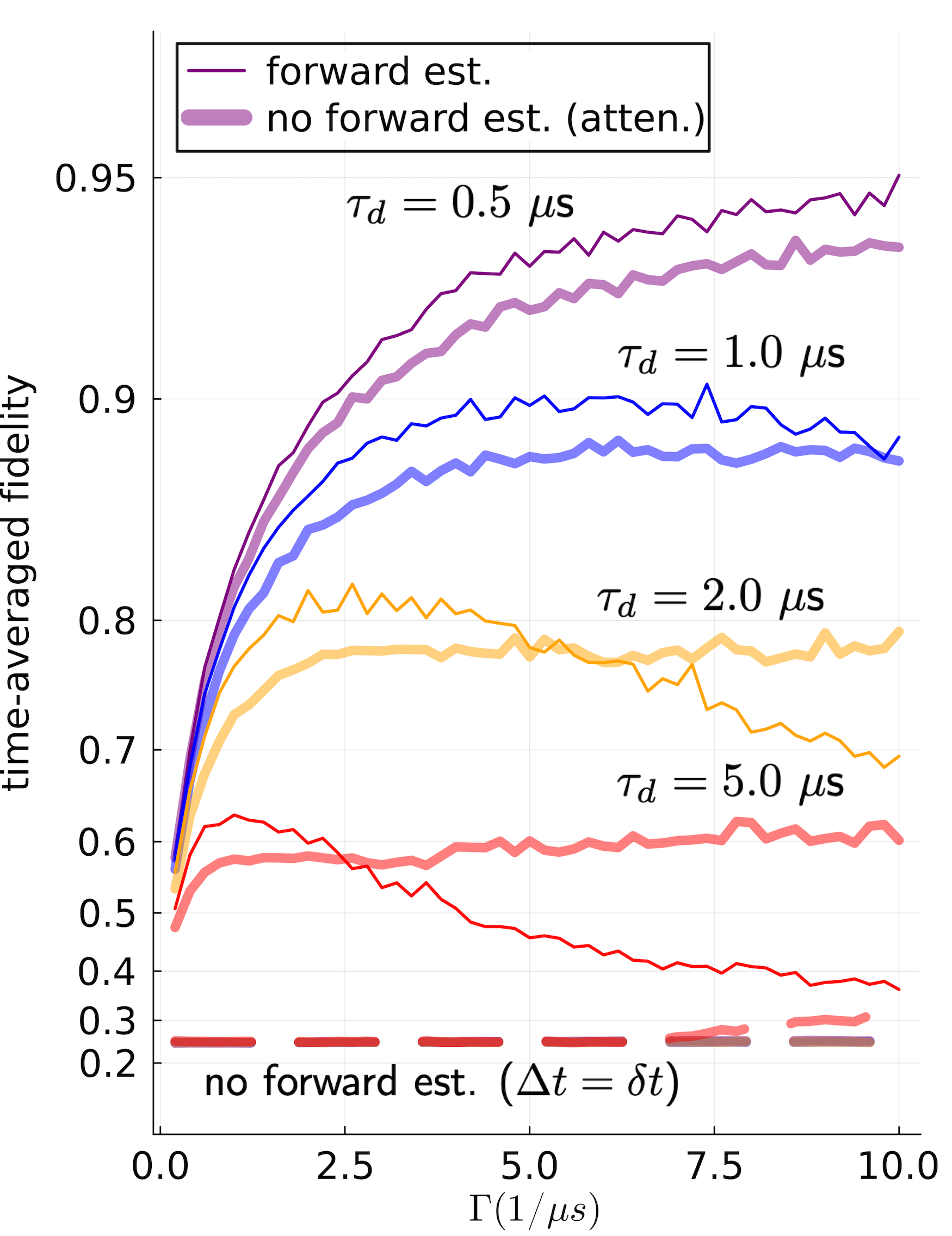}
	\caption{Ensemble fidelities averaged from $100$ to $150~\mu$s are compared on a log scale for different measurement rates $\Gamma$ (x-axis), loop delay times $\tau_d$ (color) and state estimation strategies (opaque vs. translucent), for the fluctuators plus $T_1$ noise environment with $\Gam_2 = 1/50.0$ $\mu$s, $\eta = 0.5$, $\delta t = $ 10 ns. When control does not take time delay into account, fidelity is close to the purely mixed case $\mathcal{F}=0.25$ (dashed lines). Forward estimation results in higher fidelities for all feedback loop delays $\tau_d$ after optimizing over the control parameter $\Gam$, with typical infidelity reduction by $\sim 20 \%$, demonstrating the optimality of the protocol.}
	\label{fig:estimation-strategy-comparison}
\end{figure}

\subsection{Optimizing Control Parameters}
\label{subsec:optimizingControl}

Fig. \ref{fig:heatmaps} shows asymptotic fidelities at $t = 150 ~\mu$s, again applying the best strategy for each noise environment. The left panels, Fig. \ref{fig:heatmaps}a-c, sweep the best strategy's main control parameter ($\Delta$ for $\Delta_\text{d.d.}$ and $\Gamma$ for $\Omega_\text{opt}(\rho)$) vs. the overall dephasing rate $\Gam_2$. As noted previously, for pure fluctuators or white noise (Figs. \ref{fig:heatmaps}a, c), dephasing is entirely due to fluctuations ($\Gam_2 = \Gam_{fl}$), whereas for mixed fluctuators and $T_1$ (Fig. \ref{fig:heatmaps}b), dephasing is evenly split between fluctuations and $T_1$ contributions ($1/2T_1 = \Gam_{fl} = \Gam_2/2$). Fixed parameters are $\tau_d = 0.5 ~\mu$s and $\Gam = 1/\mu$s for fluctuators. The right panels, Fig. \ref{fig:heatmaps}d-f, sweep the measurement rate $\Gam$ vs. the feedback loop delay $\tau_d$ for each noise environment, with fixed $\Gam_2 = 1/50.0\mu$s, $\eta = 0.5$, and $\Delta_0/2\pi = 25$ MHz for fluctuators. 

Similarly to Fig. \ref{fig:fidelity-timeseries}, we observe that the pure fluctuators noise environment shows the highest stabilization fidelities, with $\mathcal{F}(t_f) > 0.98$ for any $\Gam_2 \in [1/100.0\mu\text{s},~ 1/10.0\mu\text{s}]$ after optimizing counterrotating drive strength $\Delta_0$. Similarly high fidelities are possible for all loop delays $\tau_d \in [0.1~\mu\text{s}, 5.0~\mu\text{s}]$ after optimizing over $\Gam$. 
 In particular, we highlight an optimal fidelity of $\mathcal{F} = 0.9999$ with control parameters $\Gam = 5.76 / \mu$s and $\Delta_0/2\pi = 25.0$ MHz, operating at loop delay $\tau_d = 0.52~\mu$s and dephasing rate $\Gam_2 = 1/50.0\mu$s. While fluctuators plus $T_1$ and white noise environments demonstrate lower fidelities, we note that optimizing over measurement strength $\Gam$ still leads to high fidelities for realistic experimental parameters: using measurement strength  $\Gam = 10.0/\mu$s while operating at $\Gam_2 = 1/50.0\mu$s and $\tau_d = 0.52~\mu$s, optimal fidelities achieved are $\mathcal{F} = 0.95$ (fluctuators noise plus $T_1$) and $\mathcal{F} = 0.92$ (white noise). This feedback delay is achievable with modern FPGA-based controllers, which we have tested and verified with placeholder calculations on a Quantum Machines OPX controller; all other parameters are well within the current state of the art.

\subsection{Optimality of Forward Estimation}
\label{subsec:forwardTests}
We note that accounting for time delay in our state estimate is essential to the success of these protocols. If the raw time-delayed state is used in Eqs.~\eqref{eq:delta-opt}-\eqref{eq:omega-opt}, we see only instability as the controller ``overcorrects'', driving the system when previously-emitted tones would already rotate to the target state. Our forward estimation protocol addresses this issue. However, the behavior is reminiscent of a classical feedback controller with a feedback gain that is too large, which suggests a simpler protocol: turn down the strength of the feedback drives. We tested this deliberate weakening of the feedback drives such that $\Omega = \Omega_\text{opt} (\delta t / \Delta t)$. In this simpler protocol, the non-forward-estimated (time-delayed) state is used as the state estimate, and $\Delta t$ is a free parameter whose optimal value depends on other system timescales. 

Three estimation strategies are compared in Fig. \ref{fig:estimation-strategy-comparison} for different loop delays $\tau_d$ and measurement rates $\Gamma$: no forward estimation with $\Delta t = \delta t$ (no attenuation), no forward estimation with the attenuation $\Delta t / \delta t$ optimized for each $(\tau_d, ~\Gamma)$ configuration, and forward estimation with $\Delta t = \delta t$. The first case leads to complete mixing of the state, resulting in a time-averaged fidelity of $0.25$. This shows that some correction for feedback delay is essential. The second case --- no forward estimation, but with attenuated drive ---provides reliable stabilization, but the asymptotic fidelity achievable (averaged over the interval $[100,150]~\mu$s to reduce noise) is generally lower than for the forward estimation. In particular, we note that tuning over the controllable parameter $\Gam$ indicates forward-estimation as optimal for all $\tau_d$ tested, reducing infidelity by $\sim 20 \%$ compared to the attenuated drive protocol. Thus, due to the low computational overhead of forward-estimating the state, we selected forward estimation as the optimal estimation strategy for the data presented in Figs. \ref{fig:fidelity-timeseries}-\ref{fig:heatmaps}, where for simplicity $\Delta t = \delta t$ was used for all forward-estimation simulations; optimizing over the feedback attenuation while still using forward estimation could further improve fidelity.

The attenuation-without-forward-estimation strategy is slightly simpler and so may be preferred in an experiment where computational overhead is the limiting factor, at the cost of lower fidelity. In a similar attempt at simplification, we also explored a version of the protocol with quasi-linear feedback, where the feedback signal at time $t$ is linearly proportional to the measurement signal $r(t)$ (see Appendix \ref{app:linearFeedback}). The proportionality constant derived depends on the prior state estimate, so it is not true linear feedback. However, while the linear feedback is qualitatively similar to the optimal feedback for stable trajectories, on average it does not produce robust stabilization, which we attribute again to the finite time delay (see Appendix \ref{subsec:correspondence}).

\section{Conclusion and Discussion}
We have presented and simulated a protocol that reliably generates and stabilizes an entangled state of two noninteracting qubits under realistic experimental conditions with dephasing from various realistic noise environments. More generally, in the limit of measurement strength much larger than other noise scales, our protocol produces a steady state that is self-healing for arbitrary error operators and without the use of ancillas. The success hinges on two key innovations of our protocol. First, we use a drive that both provides dynamical decoupling and transfers population out of an undesired fixed point of the dynamics, working concurrently with the feedback drive. Second, we develop a forward-estimation protocol for the feedback controller that compensates for loop delay, which can be used to improve other feedback control protocols. Together these provide evidence that continuous feedback control can achieve capabilities beyond what can be replicated with discrete (projective) measurement and gates, even in the presence of realistic nonidealities.

We note that, while our simulations take into account many experimental non-idealities, there are still approximations we have made. We treat the qubits as true two-level systems, while superconducting qubits are often weakly anharmonic. We note that the bandwidth of our drives ($2/\delta t = 50$ MHz) is small compared to typical anharmonicity of a transmon ($\approx 200$ MHz), and so this assumption appears justified. We ignore stray qubit-qubit coupling, which can be eliminated in many architectures via multi-path coupling (e.g., \cite{yanTunableCouplingScheme2018a}). We ignore classical control crosstalk, which can be eliminated through proper calibration. We also ignore the effects of finite readout cavity bandwidth (i.e., we make the ``bad-cavity'' approximation of Markovianity) and the effects of an improperly tuned measurement that can distinguish  $\ket{01}$ from $\ket{10}$ (i.e., we assume a perfect half-parity measurement). These last assumptions are the most difficult to achieve experimentally. Future work could examine protocol performance when they are relaxed, for example by modeling cavity evolution directly as a truncated Fock space and including asymmetric measurement rates for $\ket{01}$ from $\ket{10}$.

Extensions of our work could optimize tradeoffs between locally optimal feedback and dynamical decoupling based on noise spectra; optimize the forward estimation protocol; and explore the impact of non-Markovian models of relaxation. Furthermore, our work can be generalized to protocols stabilizing other entangled states, including states with higher numbers of qubits. The generalization requires three components: (1) a joint measurement that has the target entangled state as an eigenstate; (2) combined single-qubit drives that rotate toward this target state from an entangled state that is not a measurement eigenstate (in our case the co-rotating drive); (3) combined single-qubit drives that have the target state as an eigenstate, but rotate all other degenerate measurement eigenstates to non-eigenstate entangled states (in our case the counter-rotating drive). We leave the rigorous general analysis of this class of protocols to future work, but note that it would be fruitful to compare protocols obtained in this way to protocols obtained via more general reinforcement learning or gradient descent methods \cite{porottiDeepReinforcementLearning2022,porottiGradientAscentPulseEngineering2023}. 
  Applying machine-learning methods to finding optimal entanglement-generation protocols could guide the choice of generalized measurements, while identifying specific features of the protocols (e.g. the three components previously mentioned) could restrict the optimization space to reduce computational overhead of machine-learning.
  We anticipate that our results will guide future studies of analog continuous measurement feedback protocols and provide an experimental roadmap for implementing fast digital processing while compensating for feedback delay.


\textbf{Acknowledgements}.---%
ELF thanks the members of QNL for facilitating this collaboration. SG thanks Arjendu Pattanayak for helpful conversations about the system-estimate distinction in quantum feedback control and  Vinay Tripathi for discussions of simulating $1/f$ noise using fluctuators. SG also acknowledges funding from the Graduate Fellowships for STEM Diversity. This work was supported by the ONR under N00014-21-1-2688, by Research Corp under Cottrell Scholarship 27550, by NSF-BSF under Grant Award No. 1915015, and by ARO/LPS under Grant Award No. W911NF-22-1-0258. 




\appendix
 
\section{Modeling dephasing}
 	\label{sec:simulating-dephasing}

 	We used three different simulation methods to model dephasing:
 	
 	\begin{enumerate}
 		\item \textbf{Lindblad}: We use time-discrete and completely positive Kraus maps to implement Markovian dephasing or decay for individual trajectories when more detailed modeling is not needed.
 		\item \textbf{Fluctuator noise}: As a more physically realistic dephasing model, we simulate a bath of uncorrelated two-level fluctuators that are dispersively coupled to each qubit. The random switches of the fluctuators make the qubit frequencies stochastic processes with $1/f$ noise spectra, leading to non-exponential decoherence when ensemble-averaged.
 		\item \textbf{Markovian white noise}: For comparison to the fluctuator model, we also simulate a white noise fluctuation model where the qubit frequencies become Markovian stochastic processes, leading to exponential decoherence of Lindblad type when ensemble-averaged.
 	\end{enumerate}
 	
 	We now summarize the theoretical background and implementation for each model and show numerically simulated examples.

   	\subsection{Lindblad decoherence}
 	We simulate average-sense decoherence using the ``jump / no-jump'' method for minimally unraveling continuous Lindblad evolution into discrete-time completely positive maps (e.g., see Eqs.~(9-10) of \cite{khezriQubitMeasurementError2015}):
  \begin{subequations}
      \begin{align}
 			\rho(t) \mapsto \rho(t + dt) &= \hat J_0 \rho(t) \hat J_0^\dagger + \sum_{k=1}^n (\gam_k  dt ) \hat J_k \rho(t) \hat J_k^\dagger, 
    \\
 			\hat J_0 &=  \sqrt{\id -  \sum_{k=1}^n (\gam_k dt) \hat J_k^\dagger \hat J_k}.
 		\end{align}
 		\label{eq:lindblad-map}
  \end{subequations}
 		
 	This unraveling models rare events that occur through one of $n$ available channels, modeled by Kraus operators $\br{\hat J_k}_{k=1}^n$ with rates $\gam_k$, such that the probability of one event in a short duration $dt$ approximates $\gamma_k dt$. The corresponding non-event Kraus operator $\hat J_0$ follows from the POVM normalization condition $\id = \hat{J}_0^2 + (\gamma_k dt)\hat{J}_k^\dagger \hat{J}_k$. The total update then averages over all $n+1$ event possibilities assuming a lack of knowledge regarding which event occurred. Using these time-discrete Kraus maps guarantees that the evolution of each time step is completely positive, so is less error-prone than na\"{i}vely applying generic numerical integration methods to the Lindblad differential equation. The usual continuous Lindblad equation can be recovered by taking the limit $dt\to 0$ of this discrete update, with the anti-commutator term arising from non-events.
  
  While we emphasize fluctuator and white noise models of dephasing as the more physically realistic models for each sampled evolution trajectory, this pre-averaged Lindblad decoherence model is an efficient way to model the expected ensemble-averaged dephasing evolution by applying it to each individual trajectory instead, acknowledging the absence of more detailed knowledge of the fluctuation mechanism for that trajectory. In particular, we use this crude ensemble-averaged model of the dephasing in projected forward-time-estimations of the system state to compensate for signal propagation delay and maximize real-time feedback performance. For this purpose, single-qubit environmental dephasing using Eq.~\eqref{eq:lindblad-map} uses the phase-jump operator $\hat J = \hat \sigma_z/\sqrt2$ with rate $\Gam_2$. Similarly, residual ensemble-dephasing due to inefficient measurement (see Section~\ref{sec:efficiency} below) uses the jump operator $\hat J = \hat N/\sqrt{2}$ with observable $\hat N = (\hat\sigma_z^{(1)} + \hat\sigma_z^{(2)})/2$ and rate $(1-\eta)\Gam$, where $\Gam$ is the ensemble-averaged measurement-dephasing rate, $\eta\in[0,1]$ is the measurement efficiency, and $\hat N$ is the half-parity operator being measured.

 	\subsection{Non-Markovian fluctuator noise}

    The fluctuators approach to simulating dephasing postulates a set of $N$ two-level systems, each of which dispersively interacts with the qubit to create classical telegraph noise. The two-level systems randomly switch between their two states in a Poisson process with characteristic rates $\{\gamma_i\}_{i=1}^N$. The rates are distributed log-uniformly between minimum  and maximum frequencies ($\gam_1, \gam_N$), i.e.,
    \begin{subequations}
        \begin{align}
            \label{eq:log-distributed}
        &\log\frac{\gamma_i}{\gamma_1} = \frac{i-1}{N-1} \log\frac{\gamma_N}{\gamma_1}\qquad \\
        &\implies \qquad \gamma_i = \gamma_1\,\left(\frac{\gamma_N}{\gamma_1}\right)^{(i-1)/(N-1)}.
        \end{align}
    \end{subequations}
    Because this distribution is exponentially weighted towards low frequencies, the resulting fluctuations have nontrivial temporal correlations, leading to non-Markovian frequency noise that goes beyond simple Markovian (e.g., Lindblad) models of dephasing. This behavior better represents the experimentally observed correlations caused by two-level systems on-chip, which are a common source of dephasing in superconducting qubit systems. Simulating dephasing with fluctuator noise in this way generates a $(1/f)^\alpha$ power spectral density (PSD) 
\cite{paladinoNoiseImplicationsSolidstate2014b,chenHOQSTHamiltonianOpen2022,tripathiModelingLowHighfrequency2023,yipOpensystemModelingQuantum2021}, which is in qualitative agreement with experimentally measured PSD, particularly in cases where the device is charge- or flux-noise sensitive \cite{paladinoNoiseImplicationsSolidstate2014b,bylanderNoiseSpectroscopyDynamical2011,trappenDecoherenceTunableCapacitively2023,yoshiharaDecoherenceFluxQubits2006}. To implement the entanglement-generation experiment we consider in the main text, a flux-tunable qubit design may be necessary to engineer the desired joint measurement---as in \cite{livingstonExperimentalDemonstrationContinuous2022}---in which case we should expect $1/f$ flux noise to be relevant.

\begin{figure}
 		\centering
 		\subfloat[Time series comparison]{\includegraphics[width=0.9\linewidth]{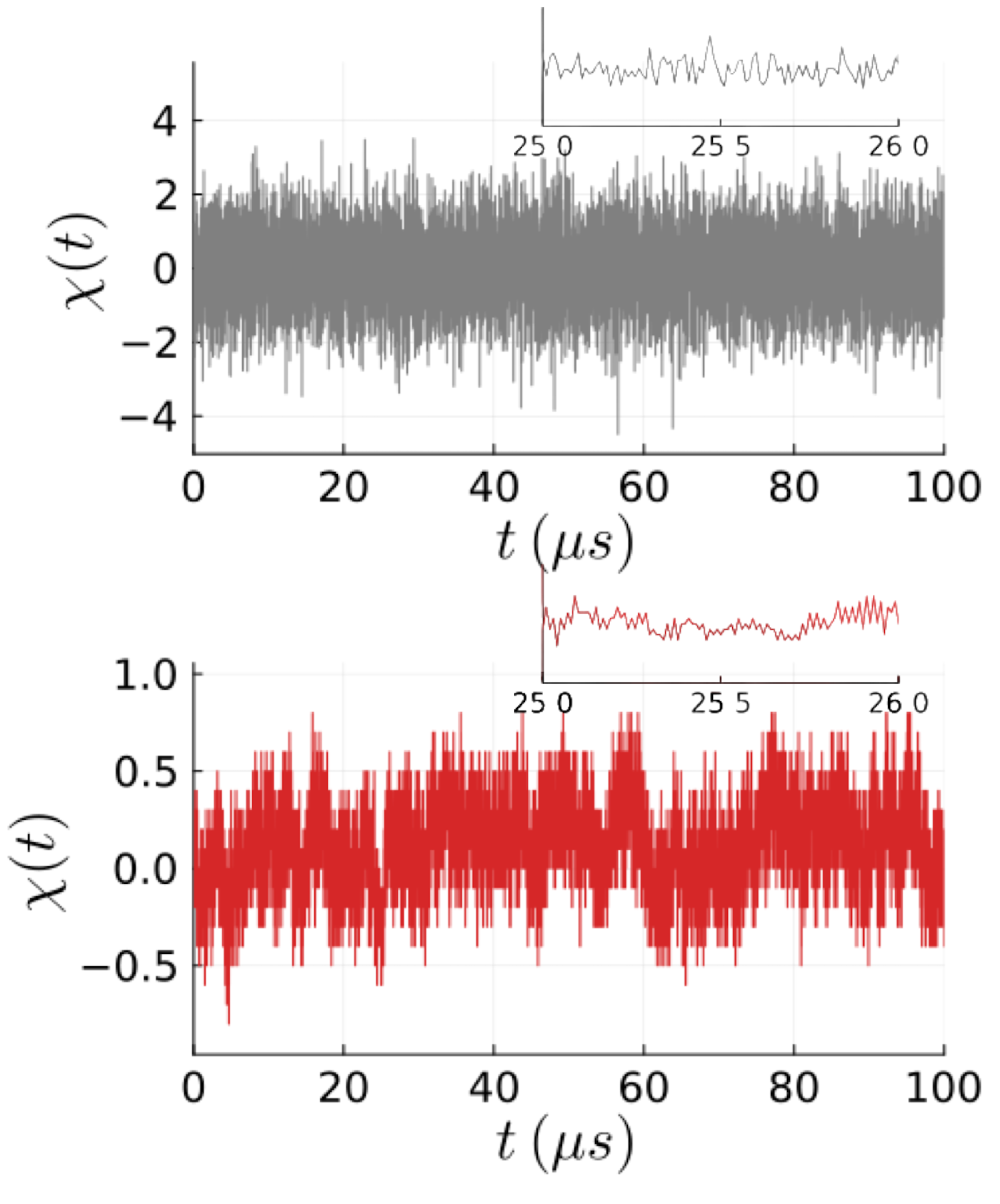}  
 			\label{fig:noise-spectra-comparison-timeseries}} \\
 		\subfloat[Power spectral density comparison]{\includegraphics[width=.52\textwidth]{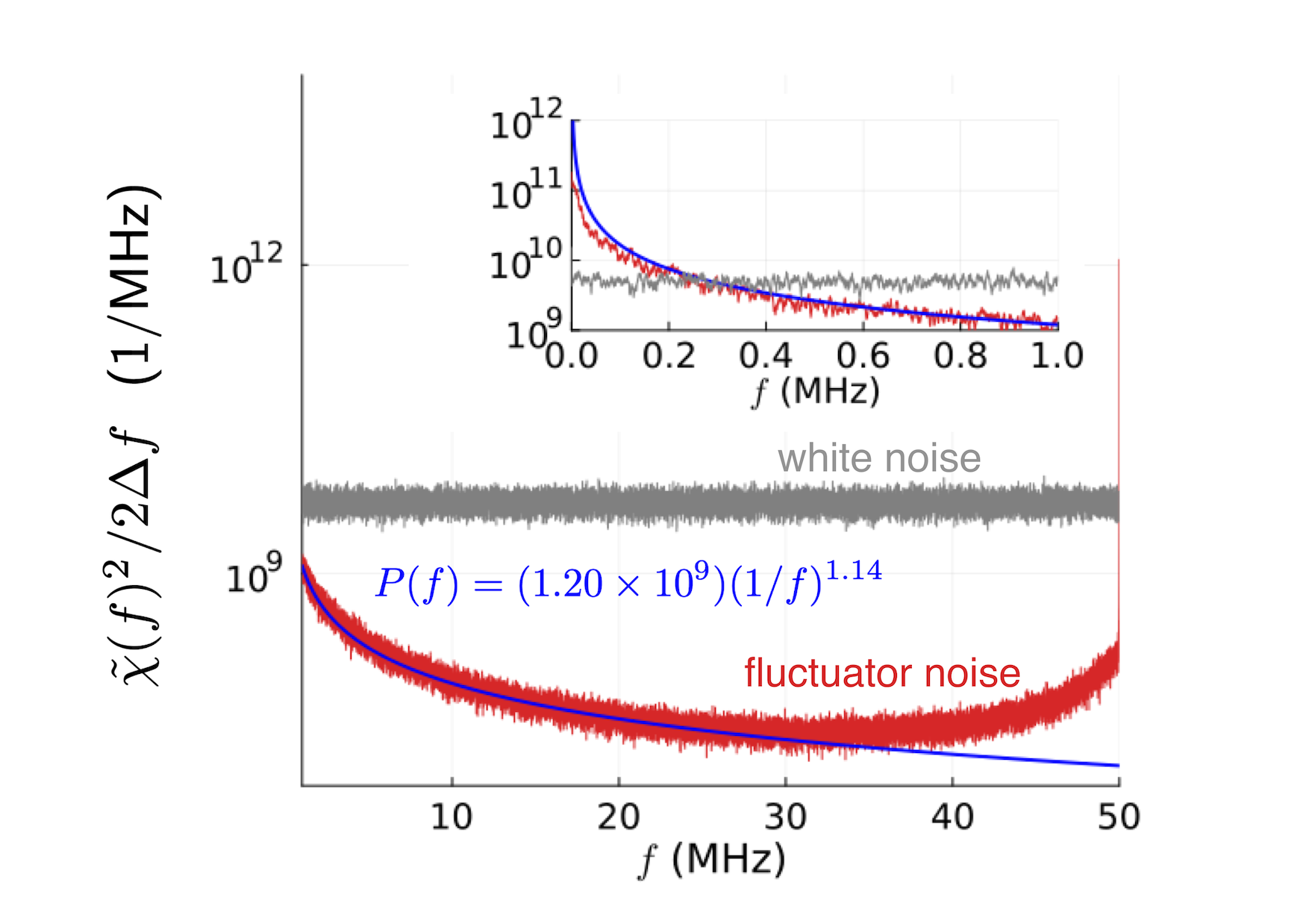}  
 			\label{fig:noise-spectra-comparison-PSD}}
 		\caption{Comparison of the temporal and spectral characteristics of white noise and fluctuator noise. (a) The white noise process (top) is stationary and self-similar at different timescales, as shown in the inset comparison, while the fluctuator noise process (bottom) is temporally correlated and thus more structured. (b) The PSD of white noise is frequency-independent, while the PSD of our sampled fluctuator noise has approximately $1/f$ frequency-dependence (blue fit curve) between 0.1 MHz and 25 MHz, with roll-off into a flat spectrum at both high and low frequencies. Note that the power spectra have been smoothed with a moving window-average of width $50 \Delta f = 1$ MHz for visual clarity.}
 		\label{fig:noise-spectra-comparison}
 	\end{figure}

    For the simulations in the main text, we generate each stochastic telegraph noise realization with the following procedure:
 	\begin{enumerate}
 		\item We define $N = 20$ frequencies $\{\gam_i = 2\pi\,f_i \}$ distributed
 		log-uniformly between $(f_1, f_N) = (5 \text{ kHz}, 50 \text{ MHz})$ according to Eq.~\eqref{eq:log-distributed}.
 		
 		\item For each $\gam_i$, we define a fluctuator to be a classical stochastic process $s_i(t)$ that takes the binary values $-1$ or $1$ for all $t\in[0,t_f]$ up to the simulation duration $t_f$. Each fluctuator randomly switches between its two values as a Poisson process with characteristic switching rate $\gam_i$, starting from an equiprobably random initial state $s_i(0) = \pm 1$.
 		
 		For efficient simulation, we use the fact that the durations between random switches of a Poisson process are exponentially distributed, $T_j^{(i)} \sim \text{Exp}(\gam_i)$, i.e., with probability density function $p(T_j^{(i)}) = \gam_i \exp(-\gam_i T^{(i)}_j)$. Thus, we first sample the random durations between switches until the total duration, $\sum_{j=1} T_j^{(i)}$, exceeds the desired simulation duration $t_f$. We then construct the stochastic process $s_i(t)$ as a piecewise-constant function that flips sign after each duration $T^{(i)}_j$ in the sampled sequence, starting from $s_i(0)$.
   
 		
 		\item Finally, we construct the the total stochastic process that affect each qubit frequency by averaging the effect of all $N = 20$ fluctuators. The resulting stochastic process $\chi(t) = (1/N) \sum_{i=1}^{N} s_i(t)$ has zero mean and a PSD with ${}\sim 1/f$ dependence, as verified in Fig.~\ref{fig:noise-spectra-comparison}.
 	\end{enumerate}

 	
  The simulated fluctuator envelopes modulate the qubit frequencies in the effective Hamiltonian,
 	\begin{equation}
 		\hat H_{\rm fluct}(t) = \frac{\om}{2} \chi_1(t)\, \hat \sigma_z^1 + \frac{\om}{2} \chi_2(t)\, \hat \sigma_z^2,
 		\label{eq:fluctuator-ham}
 	\end{equation}
  where $\chi_{1,2}(t)$ are independent fluctuator processes generated with the above procedure, while $\om$ is a (constant) characteristic frequency that sets the strength of the dispersive frequency fluctuations. 
  
  Notably, after ensemble averaging, these frequency fluctuations produce a characteristic decay envelope that is distinct from the purely exponential decay of Markovian (Lindblad) dephasing models, as can be seen in Fig.~\ref{fig:noise-spectra-comparison}, with quadratic decay at small times rather than linear decay. 
  However, at longer times the fluctuator coherence decay more closely approximates exponential decay, with the fluctuation strength $\om$ indirectly setting a characteristic decay rate $\Gamma_2$ that can be compared to the exponential decay rate of Markovian dephasing models. We describe how to calibrate the scaling frequency $\om$ in to achieve a desired $\Gamma_2$ in Sec. \ref{subsec:noise-equivalency}.

 \subsection{Markovian white noise}
 \label{subsec:white-noise}
 For comparison with the non-Markovian fluctuator model, we also simulate explicitly Markovian (white noise) fluctuations that produce purely exponential decay of the ensemble-averaged coherence. Using the same Hamiltonian as Eq.~\eqref{eq:fluctuator-ham}, each stochastic envelope $\chi_k$ is instead sampled at discrete times $t_\ell = t_0 + \ell\,dt$ from a zero mean Gaussian distribution, $\chi_k(t_\ell) \sim \text{Norm}(0,\tau/dt)$ with variance $\tau/dt$ characterized by a timescale $\tau$ and the time discretization $dt$. The resulting envelope functions $\chi_k(t)$ then are Markovian random processes with a piecewise constant structure that limit to white noise processes satisfying $\langle \chi_k(t)\chi_k(t')\rangle = \tau\,\delta(t-t')$ in the time-continuum limit $dt\to 0$. The exponential decay rate of the ensemble averaged Bloch radius in the $x$-$y$ plane is then,
 \begin{subequations}
  \begin{align}
      \Gamma_2 &= \int_{t_0}^t\left\langle [\omega\chi_k(t)][\omega\chi_k(t')] \right\rangle dt' = \frac{1}{2}\omega^2\tau, \\ & \implies \qquad \omega = \sqrt{\frac{2\Gamma_2}{\tau}}.
  \end{align}
  \label{eq:white-dephasing-analytic}
 \end{subequations}
 We numerically confirm this scaling of the decay rate with $\omega$ in Fig.~\ref{fig:white-fit} of the next section. For simplicity in the simulations, we choose $\tau = dt$ to be the same as the time step duration, making the sampled $\chi_k$ have unit variance.

 	\subsection{Calibrating the fluctuation strength for a target dephasing rate}
 	\label{subsec:noise-equivalency}
 	To fairly compare dephasing timescales produced by the different noise models, we numerically calibrate the dependence between the noise strength parameter $\om$ in the Hamiltonian in Eq.~\eqref{eq:fluctuator-ham} and a rate $\Gam_2$ that characterizes the dephasing envelope. 
 Importantly, the strength $\omega$ is not directly measurable experimentally, while the ensemble-averaged decay profile of the coherence can be observed in a Ramsey experiment. Calibrating $\omega$ as a function of the characteristic rate $\Gam_2$ thus better connects our simulated noise models to experimental observation. Moreover, matching effective decay rates in the absence of feedback is critically important to meaningfully compare and assess the performance of each feedback protocol in the main text.

A subtle complication with establishing a meaningful rate comparison is that while exponential decay is memory-less with a constant decay rate $\Gam_2$, fluctuator noise has a temporally structured decay that is is quadratic for short times and exponential at late times, so any fit to purely exponential decay is problematic. This non-Markovian behavior arises from the exponential memory kernel of the fluctuator noise time correlations and is precisely what makes it possible to dynamically decouple from short-time low-frequency fluctuator noise experimentally (e.g., using a spin echo mid-gate). Nevertheless, we can still identify a characteristic rate $\Gamma_2$ that sets the time scale of the total decay profile and reduces to the exponential decay rate in the Markovian limit, which is sufficiently close for the purposes of the main text. Most importantly, we confirm that the feedback protocol fidelity results in the main text are not artifacts of any miscalibration of our fluctuator models, but are instead fundamental to the choice of time-dependent feedback.
  
 	Since we can crudely think of the fluctuator noise spectrum as a mixture of $1/f$ and white noise, we can postulate a mean decay profile along a ``stretched exponential'' that interpolates between the two. To calibrate $\omega$, we simulate Ramsey experiments as follows: we initialize ensembles of single-qubit $\ket{+x}$ states undergoing fluctuator or white noise, sweep the parameter $\om$ over the range $[0.25,\,1]$ /$\mu$s for fluctuators noise and $[1,\,5]$ /$\mu$s for white noise, extract the ensemble-averaged Bloch coordinate $x$ as a function of time, and fit the resulting decay curves to a stretched-exponential function,
  \begin{equation}
      x(t) = \exp\bigp{-(\Gam_2 t)^\beta}, 
      \label{eq:stretched-exp}
  \end{equation}
  characterized by both a rate $\Gam_2$ and a dimensionless power $\beta$. The power $\beta = 1$ corresponds to purely exponential decay, while the short-time decay anticipated for $1/f$ noise should be quadratic with $\beta = 2$, suggesting that an average fit for $\beta$ will be somewhere between 1 and 2 for the full fluctuator noise decay profile. Eq.~\eqref{eq:stretched-exp} is a common experimental choice for fitting non-exponential decay with a single rate parameter and fits the broad structure of our simulated decay curves reasonably well. Moreover, the extracted rates $\Gam_2$ for non-exponential decay are also very close to the rates obtained from na\"ive exponential fits to same data, which supports the use of the obtained $\Gam_2$ as a reasonable comparison to purely exponential decay rates.

  	\begin{figure}
 	\centering
 	\subfloat[Fluctuator Ramsey decay, varying $\om\in (0.25,1.0)$ rad/s]{
 		\includegraphics[width=\linewidth]{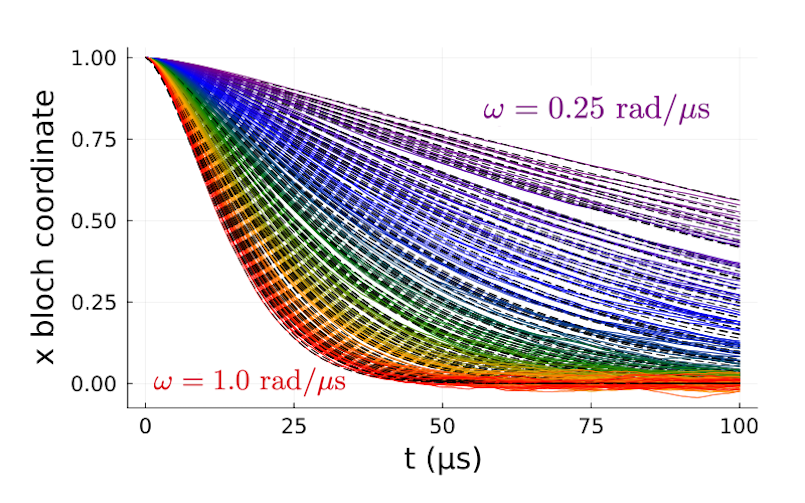}  
 		\label{fig:fluctuators-fit-timeseries}} \quad \\
 	  	\subfloat[Fluctuator fit parameter $\Gam_2 / \gamma_\text{geom}$ vs. varying $\om / \gamma_\text{geom}$.]{\includegraphics[width=\linewidth]{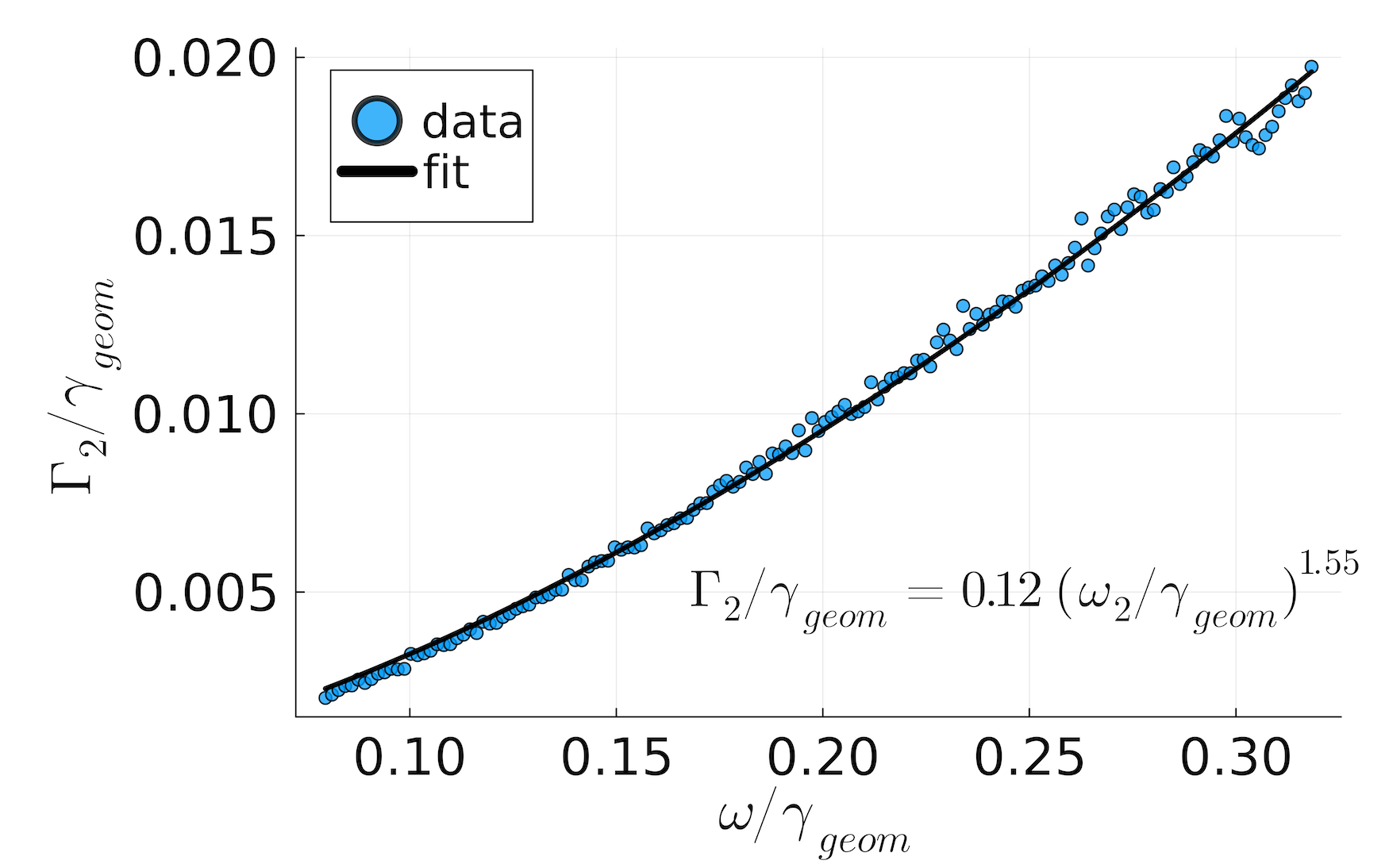}  
 	    \label{fig:fluctuators-fit}} 
      
  	\subfloat[Fluctuator fit parameter $\beta$ vs. varying $\om$.]{
 	    \includegraphics[width=\linewidth]{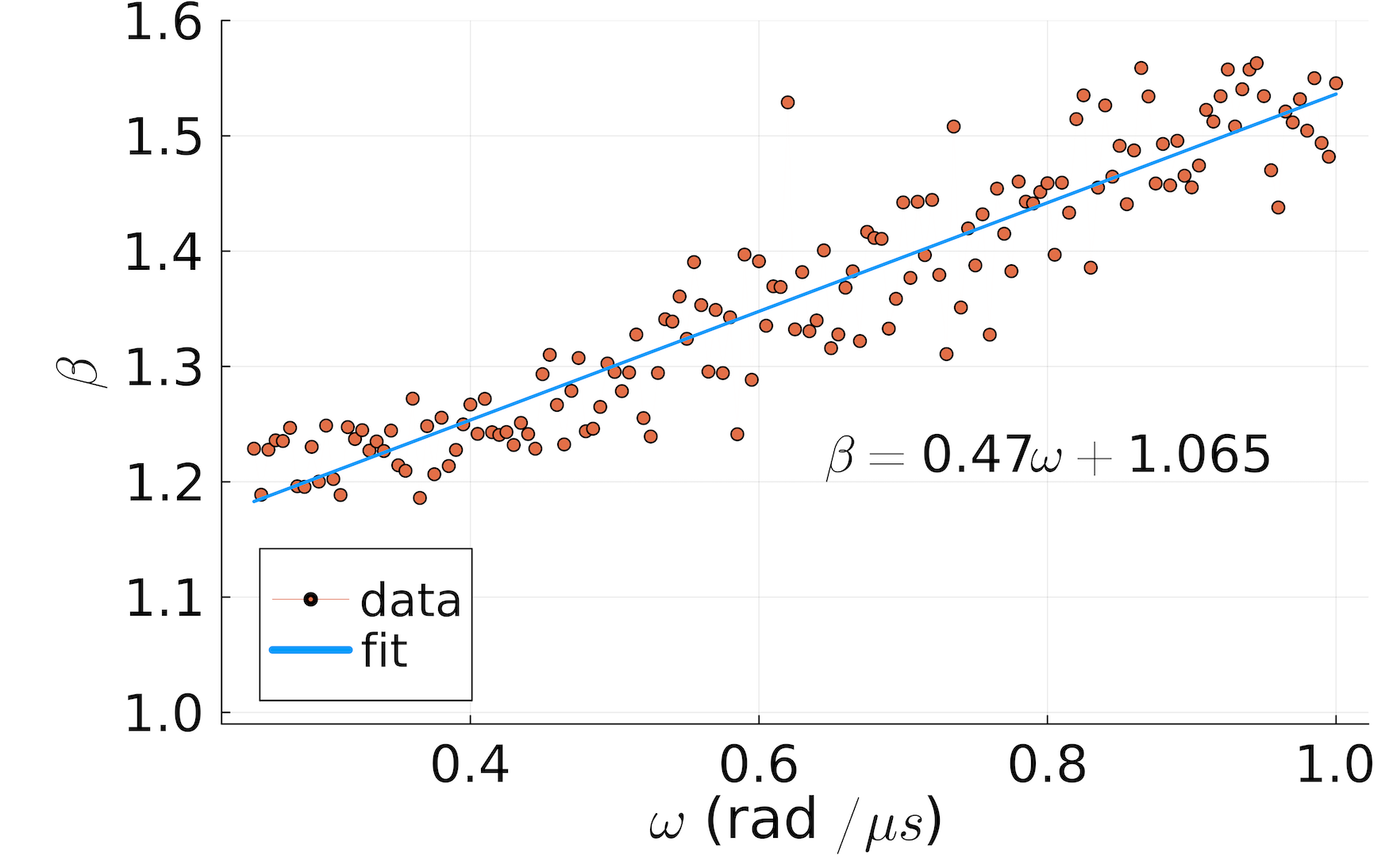}  
 	    \label{fig:fluctuators-beta-fit}} 
      
 	\caption{\textbf{(a)} Ramsey decay curves for fluctuators with different choices of frequency drift amplitude $\omega$, fit to a two-parameter stretched exponential decay, $x(t;\,\Gamma_2,\beta) =\exp(-(\Gam_2 t)^\beta)$. \textbf{(b)} Relation between drift amplitudes $\omega$ and the dephasing rate parameter $\Gam_2$, which is fit to a power law used to set the target dephasing rates in the main text. The frequency reference $\gam_\text{geom}=$ $\sqrt{\gamma_1\gamma_N} = (2\pi)\,500$ kHz is the geometric mean of the $N=20$ log-distributed fluctuator frequencies between $\gamma_1/2\pi = 5$ kHz and $\gamma_N/2\pi = 50$ MHz.  \textbf{(c)} Relation between drift amplitudes $\omega$ and the dephasing power parameter $\beta$.}
 	\label{fig:fluctuator-ramsey}
 \end{figure}

    	\begin{figure}
 	\centering
   	\subfloat[White noise Ramsey decay, varying $\om \in (1.0, 5.0)$ rad/s.]{
 		\includegraphics[width=\linewidth]{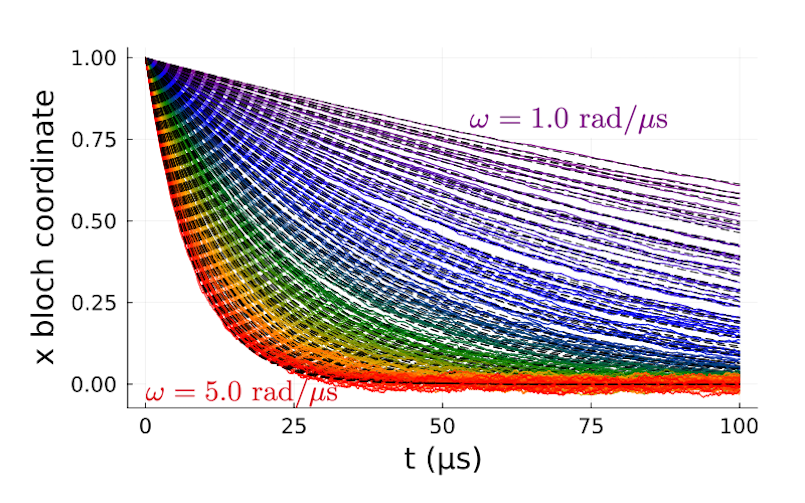}  
 		\label{fig:white-fit-timeseries}}  \quad \\
	\subfloat[White noise fit parameter $\Gam_2  ~ dt$ vs. varying $\om ~dt$.]{
 		\includegraphics[width=\linewidth]{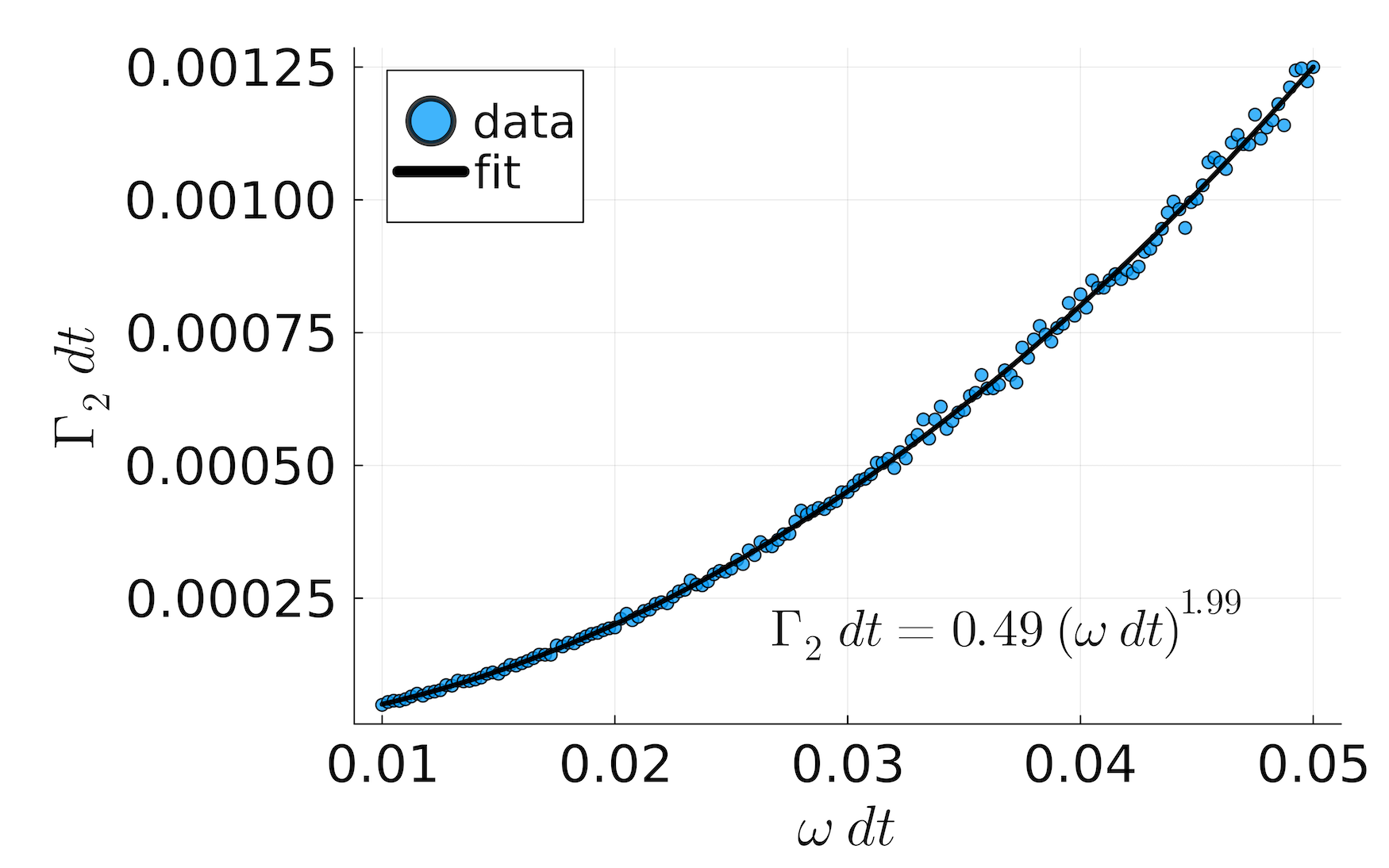}  
 		\label{fig:white-fit}}

    	\subfloat[White noise fit parameter $\beta$ vs. varying $\om$.]{
	    \includegraphics[width=\linewidth]{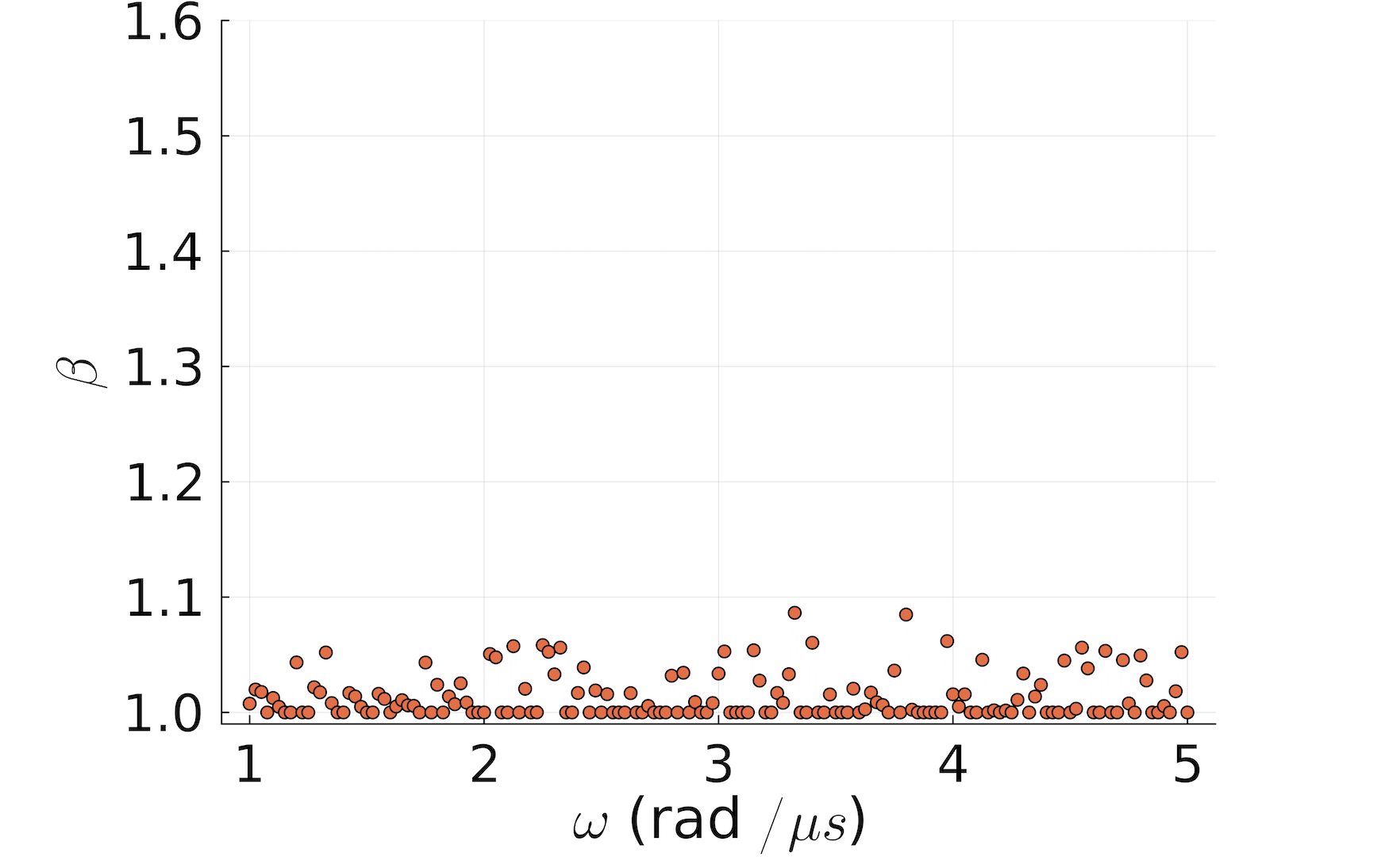}  
	    \label{fig:white-beta-fit}}

 	\caption{\textbf{(a)} Ramsey decay curves for white noise with different choices of frequency drift amplitude $\omega$, fit to a two-parameter stretched exponential decay, $x(t;\,\Gamma_2,\beta) =\exp(-(\Gam_2 t)^\beta)$. \textbf{(b)} Relation between drift amplitudes $\omega$ and the dephasing rate parameter $\Gam_2$, which is fit to a power law used to set the target dephasing rates in the main text. The white noise characteristic time $\tau = dt = 10$ ns is set to the simulation timestep duration for convenience. \textbf{(c)} Relation between drift amplitudes $\omega$ and the dephasing power parameter $\beta$. Exponential decay corresponds to the power $\beta = 1$, confirmed by the white noise fit up to small deviations due to fit imperfections.}
 	\label{fig:white-ramsey}

 \end{figure}

 For the fluctuator noise model, the stretched exponential fits shown in Fig. \ref{fig:fluctuators-fit-timeseries} determine the relationship between $\om$ and $\Gam_2$ shown in Fig. \ref{fig:fluctuators-fit}, which has the following polynomial fit:
      \begin{align}
          \frac{\Gam_2}{\gam_\text{geom}} &= 0.12 \Bigp{\frac{\omega}{\gam_{\text{geom}}}}^{1.55} &\text{(fluctuator noise).}
      \end{align}
  For this fit we choose the reference rate to be the geometric mean of the log-uniformly spaced fluctuator frequencies, $\gam_\text{geom} = \sqrt{\gam_1 \gam_N} = \exp[(1/N) \sum_i \log\gam_i] = (2\pi)(500 \text{kHz}) = \pi/\mu\text{s}$, yielding the swept parameter range $\omega/\gam_\text{geom}\in[1/4\pi,\,1/\pi]$. Other choices of frequency reference will scale the prefactor of the fit, but won't change the dependence on $\omega$. The numerical results for the prefactor (0.12) and power (1.55) generally depend on other parameters that we keep constant in our simulations, like the number of fluctuators $N$ and the number of frequency decades $d = \log_{10} (\gam_\text{max} / \gam_\text{min})$ spanned about the mean $\gam_{\text{geom}}$, but these additional dependencies are not important for the results of the main text. The relationship between $\om$ and the power $\beta$ is also shown in Fig.~\eqref{fig:fluctuators-beta-fit} and is approximately linear, $\beta(\omega) = 1.07 + 0.47(\omega/\gamma_\text{geom})$. 
  
 For the white noise fluctuation model, the stretched exponential fits shown in Fig.~\ref{fig:white-fit-timeseries} yield the relationship between $\om$ and $\Gam_2$ shown in Fig.~\ref{fig:white-fit}, with a polynomial fit,
      \begin{align}
      \label{eq:white-noise-fits}
          \Gam_2 dt &= 0.49 \bigp{\om dt}^{1.99}, &\text{(white noise)}
      \end{align}
 which confirms the dependence expected from Eq.~\eqref{eq:white-dephasing-analytic} with the noise timescale choice of $\tau = dt$. Similarly, the power $\beta$ is consistent with the exponential-decay power of 1, as shown in Eq.~\eqref{fig:white-beta-fit}.
  
  These fit equations can then be inverted to obtain the conversion from $\Gam_2$ to $\omega$ used for the simulations in the main text:
  \begin{subequations}
      \begin{align}
          \om(\Gam_2) &= 3.9\, \gam_\text{geom} \Bigp{\frac{\Gam_2}{\gam_\text{geom}}}^{0.65} &\text{(fluctuators),} \\
          \om(\Gam_2) &= \sqrt{\frac{2\Gam_2}{dt}} &\text{(white noise)}.
      \end{align}
      \label{eq:noise-conversion}
  \end{subequations}

 \subsection{Effective dephasing rate with constant dynamical decoupling Rabi drive}
 \label{subsec:effective-dephasing}
 
 In addition to calibrating the relationship between the fluctuation amplitude $\om$ in the Hamiltonian in Eq.~\eqref{eq:fluctuator-ham} and the resulting characteristic dephasing rate $\Gamma_2$, we also calibrate the effective dephasing rate $\Gamma_\text{eff}$ observed in the presence of a dynamical decoupling drive. This is applicable to the ``simultaneous d.d. and feedback'' protocol in the main text, where $\Delta(\rho) = \Delta_{\rm d.d.}$ is constant and acts to decouple the system from slow fluctuations. In the case of a single qubit undergoing dephasing with no feedback, a constant Rabi drive $\Delta\,\hat\sigma_y /2$ around the $y$-axis orthogonal to the frequency fluctuation $z$-axis $\hat \sigma_z$ will mitigate the ensemble dephasing by reducing the net time that the qubit coherence is affected by frequency fluctuations. The effective dephasing rate $0 < \Gam_\text{eff} < \Gam_2$ will be reduced in proportion to the ratio $\Delta/\Gam_2$ of decoupling and bare dephasing timescales. The effectiveness of such a decoupling technique depends on the short-time behavior of the ensemble dephasing, which is linear for exponential decay but has slower quadratic dependence for fluctuator-based dephasing. Thus, this simple dynamical decoupling technique is significantly more effective at mitigating coherence decay for fluctuator noise than for white noise.
 
 Knowing the expected effective dephasing $\Gamma_\text{eff}$ in the presence of a decoupling Rabi drive is important for our feedback protocol in the main text. The forward state estimation used to compensate for feedback delay uses an effective dephasing Lindblad term in the filter (see Sec. \ref{app:forward-estimation}), which must use an accurate effective dephasing rate that accounts for the drive. Including this effective ensemble dephasing in the forward estimation improves the state-tracking fidelity and, as a result, the feedback stabilization performance. 

\begin{figure}
    \centering
    \includegraphics[width=\linewidth]{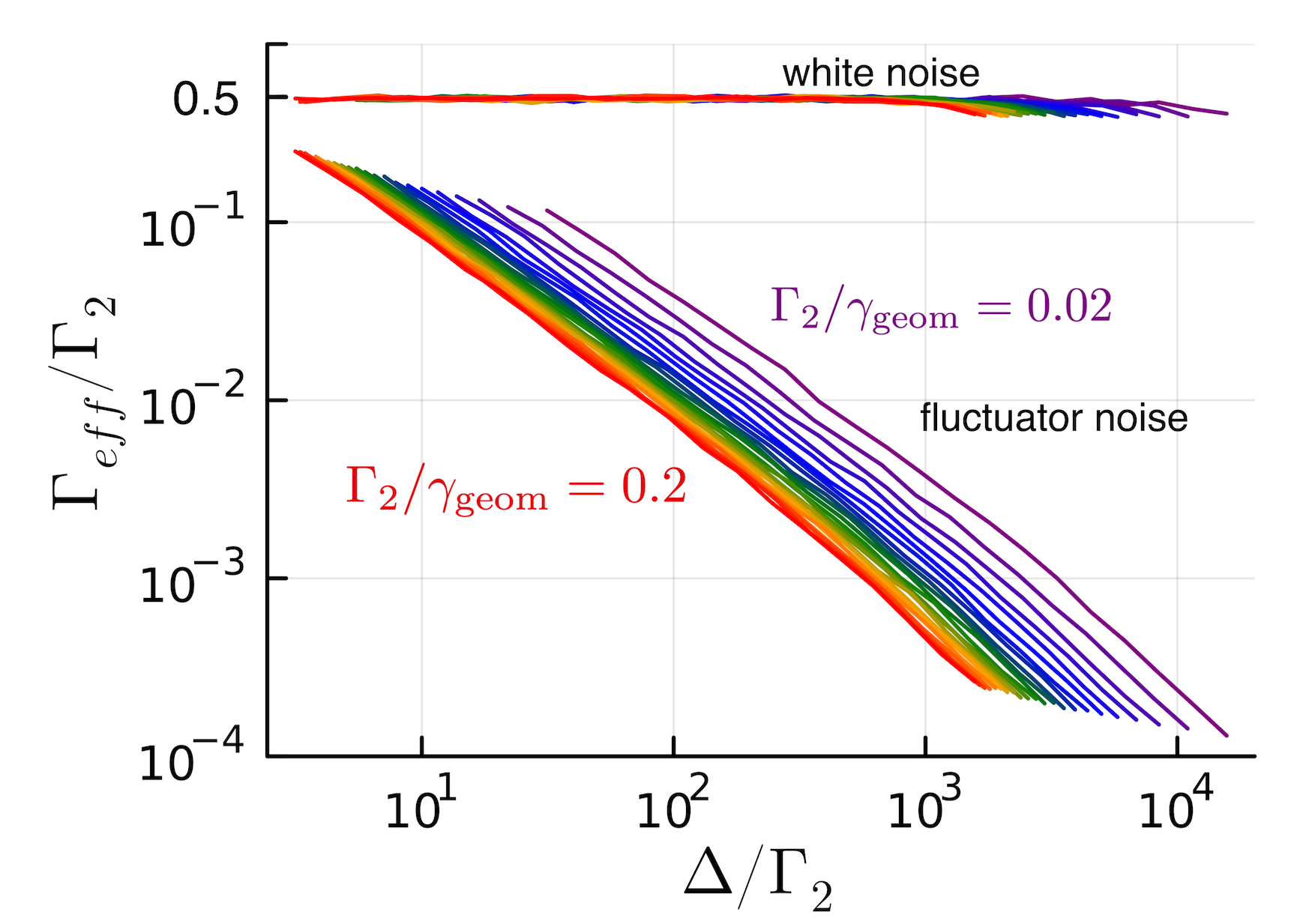}
    \caption{Ratio of effective qubit dephasing rate $\Gam_\text{eff}$ to bare rate $\Gam_2$ in the presence of a constant decoupling Rabi drive $\Delta \, \hat \sigma_y/ 2$. The qubit frequency fluctuations $\om\chi(t)\, \hat \sigma_z/2$ have the amplitude $\om$ calibrated to produce the ensemble-dephasing rate $\Gam_2$ in the absence of the decoupling drive, based on Fig. \ref{fig:fluctuators-fit} and \ref{fig:white-fit}. The small-time quadratic dependence of the fluctuator-based dephasing dramatically increases the effectiveness of the decoupling drive, with the effective dephasing rate decreasing as roughly $1/\Delta$. In contrast, for white noise the effective dephasing drops to $\Gamma_\text{eff} = \Gamma_2/2$ for any nonzero $\Delta$ due to the system state spending approximately half the time close to the dephasing-immune $z$-axis.}
    \label{fig:gamma-eff-calibration}
\end{figure}

 Fig. \ref{fig:gamma-eff-calibration} shows the effective dephasing rate $\Gam_\text{eff}$ for a sweep of decoupling Rabi amplitudes $\Delta$ and bare dephasing rates $\Gam_2$ for a single qubit. Because the qubit state vector is rotating in the $x$-$z$ plane on average, the purity (i.e., $P = \text{Tr}\rho^2 = (1 + r^2)/2$, with $r^2 = x^2 + y^2 + z^2$) is used to extract a time constant. Since the drive generally decouples from the slow noise components contributing to the non-exponential decay, a single parameter fit is sufficient,
 \begin{equation}
     P(t) = \frac{1 + \exp(-2 \Gam_\text{eff} t)}{2}.
 \end{equation}
 For $\Delta = 0$, $\Gam_\text{eff} = \Gam_2$ as expected. 
 
 For white noise, $\Delta/\Gam_2 > 1/2$ yields the asymptotic limit $\Gam_\text{eff}/\Gam_2 \to 1 / 2$, as shown in Fig.~\ref{fig:gamma-eff-calibration} and predicted from the white noise correlator $\langle \chi_k(t)\chi_k(t')\rangle = \tau\delta(t-t')$ \cite{lidarOpenQuantumSystems}. The intuition behind this limit is that under a constant Rabi drive the qubit spends about half its time near the $z$-axis, where it is immune to $\hat \sigma_z$ rotations, so dephases at half the rate. 
 
 In contrast, for fluctuators there are much larger reductions in $\Gam_\text{eff}$ as $\Delta$ is increased. As shown in Fig.~\ref{fig:gamma-eff-calibration}, for larger $\Delta/\Gam_2 \gg 1$ the rate $\Gam_\text{eff}$ decays as roughly $\Gam_2/\Delta$:
 \begin{equation}
     \Gamma_\text{eff} / \Gamma_2 = 0.09 ~( \gamma_\text{geom} / \Gamma_2)^{0.71} (\Gamma_2 / \Delta)^{1.1},
 \end{equation}
 with the $(\gamma_\text{geom} / \Gamma_2)$ prefactor accounting for the $\Gamma_2$-dependent intercepts on the log-log plot.
 
  This $1/\Delta$ behavior allows $\Gamma_\text{eff}$ to become arbitrarily small for a large enough decoupling drive. 
 For simplicity and accuracy, we use the numerical results shown in Fig.~\ref{fig:gamma-eff-calibration} to calibrate $\Gam_\text{eff}$ in the state-tracking filter used in the main text.

\section{Measurement model} 
\label{sec:bayesian}

 We use a Bayesian update model of time-stepped sequential weak measurements to describe the backaction of the joint continuous measurement on the two-qubit system. For completeness, we briefly review the essentials of this method. 
  
 In the time-continuum limit, the monitoring of an observable $\hat N$ with a homodyne readout will produce a normalized signal $r(t)$ that approximates a stochastic Wiener process
 	\begin{equation}\label{eq:wiener}
 		r(t) dt =\text{Tr}(\rho_\text{sys}(t) \hat N) dt + \sqrt{\tau} ~dW,
 	\end{equation}
with Wiener increment $dW^2 = dt$ (assuming the It\^o picture) and a characteristic measurement timescale $\tau = 1/2\Gam\eta$ related to the ensemble measurement-dephasing rate $\Gamma$ through the measurement efficiency $\eta\in[0,1]$ (see Section~\ref{sec:efficiency} below). This approximation is the foundation of stochastic Schr\"odinger equation or master equation methods for describing continuous measurement. Averaging this readout for a duration $dt$ yields a Gaussian random variate,
\begin{equation}
    r_i = \int_{t_i}^{t_i+dt}r(t')\frac{dt'}{dt} = \text{Tr}(\rho_i \hat N) + \sqrt{\frac{\tau}{dt}}~ \xi_i,
    \label{eq:discrete-r}
\end{equation}
with variance $\tau/dt$, written here in terms of its mean with additive zero-mean Gaussian noise $\xi_i$ with unit variance, i.e. $\xi_i \sim \mathcal{N}(0,1)$, which can be efficiently sampled numerically. The observable expectation value with the state $\rho_i = \rho_\text{sys}(t_i)$ approximately determines the mean value of the signal over the time bin of width $dt$. This relation also makes it clear that the measurement timescale $\tau$ represents the integration time required to achieve a unit signal-to-noise ratio.

In the laboratory, the finite bandwidth of the signal analyzer discretizes the time axis into bins $dt$ in this way, making the observed readout signal more accurately a discrete readout timeseries $\vec r = \{r_1, ..., r_i, ..., r_n\}$ with approximately Gaussian random variates. This random time series will determine the measurement backaction on the system. In other words, for a simulation with time vector $\vec t = \{t_1, t_2, ..., t_n\}$, the state evolution will be completely determined by the randomly sampled Gaussian noise $\vec \xi = \{\xi_1, \xi_2, ..., \xi_n\}$ combined with the initial condition $\rho_0$ and full knowledge of the unitary, dissipative, and measurement evolution. 
 	
 	To obtain the state trajectory $\vec \rho = \{\rho_1, \rho_2, ..., \rho_n\}$ from $\vec \xi$, we use the discrete Bayesian filter \cite{korotkovQuantumBayesianApproach2011} presented in the main text,
 	\begin{equation}
 		\rho_{i+1}= \mathcal{L}_i \frac{\hat U_i \hat M_{r_i} \rho_i \hat M_{r_i}^\dagger \hat U_i^\dagger}{\text{Tr}(\hat M_{r_i}^\dagger \hat M_{r_i} \rho_i)}. \label{eq:state-update}
 	\end{equation}
 	Here, $\mathcal{L}_i$ is a Kraus map representing total average dephasing (due to, e.g., the imperfect measurement efficiency discussed in Section~\ref{sec:efficiency}), $\hat U_i$ is a unitary representing the coherent evolution between times $t_i$ and $t_i+dt$, and $\hat M_{r_i}$ is the measurement Kraus operator that implements the measurement backaction and depends on the sampled $r_i$ as described below. The set of $\hat M_{r_i}$ over all possible $r_i$ construct a Gaussian POVM with completeness condition,
 	\begin{equation}\label{eq:povm-condition}
 		\int_{-\infty}^\infty dr ~|c(r)|^2\hat M_r^\dagger \hat M_r = \sqrt{\frac{dt}{2\pi\tau}} \int_{-\infty}^\infty dr ~ e^{-dt(r-\hat{N})^2/2\tau} = \hat I,
 	\end{equation}
  with an $r$-dependent normalization $|c(r)|^2$ factor that cancels in the state update of Eq.~\eqref{eq:state-update}.

 	\subsection{Derivation of the measurement operator}
 	To derive the measurement operator specific to our setup directly from the correspondence principle, we start from the standard form of a measurement Kraus operator for $\hat{N}=(\hat{\sigma}_z^{(1)} + \hat{\sigma}_z^{(2)})/2$ that associates the eigenspace projection operators for its three eigenvalues of $-1,0,1$ with their corresponding conditional readout likelihoods \cite{korotkovQuantumBayesianApproach2011,pattiLinearFeedbackStabilization2017},
  \begin{equation}
  \begin{aligned}
       		c(r)\hat M_r = &\sqrt{P(r|00)} ~\ket{00}\!\bra{00} + \sqrt{P(r|11)} ~\ket{11}\!\bra{11} \\ &+ \sqrt{P(r|~01,10)}~ \bigp{\ket{01}\!\bra{01} + \ket{10}\!\bra{10}},
         \end{aligned}
  \end{equation}
 	
   anticipating a state-independent factor $c(r)$ that will cancel. By construction, the POVM condition $\int dr~|c(r)|^2\,\hat M_r^\dagger \hat M_r = \hat I$ then follows from the normalization of each likelihood distribution.
  
 	For a calibrated readout $r$ averaged over a time bin $dt$, the readout likelihood distributions, 
 	\begin{subequations}
 		\begin{align}
 			P(r|00) &= \sqrt{\frac{dt}{2\pi\tau}} \exp\Bigp{-\frac{dt}{2\tau} (r - 1)^2} \\
 			P(r|11)&= \sqrt{\frac{dt}{2\pi\tau}} \exp\Bigp{-\frac{dt}{2\tau} (r + 1)^2}\\
 			P(r | 01,10) &= \sqrt{\frac{dt}{2\pi\tau}} \exp\Bigp{-\frac{dt}{2\tau} r^2},
 		\end{align}
 	\end{subequations}
 	are each Gaussian with variance $\tau/dt$ centered at an eigenvalue of $\hat{N}$ for each eigenspace, assuming the quantum computing convention $\hat{\sigma}_z = \ket{0}\!\bra{0} - \ket{1}\!\bra{1}$. Thus, 
 	\begin{subequations}
        \begin{align}
        \hat M_r &= e^{(dt/4\tau)(2r-1)} \ket{00}\!\bra{00} \\&~+ e^{-(dt/4\tau)(2r+1)} \ket{11}\!\bra{11} + \bigp{\ket{01}\!\bra{01} + \ket{10}\!\bra{10}}, \nonumber \\
            c(r) &= \left(\frac{dt}{2\pi \tau}\right)^{1/4}~e^{-(dt / 4\tau) r^2},
        \end{align}
 	\end{subequations}
        are the forms of the state-dependent Kraus operator $\hat M_r$ and the state-independent normalization coefficient $c(r)$ that will cancel in the state update. 
    Written in terms of the Bell basis, this measurement operator has the useful closed form,
    \begin{equation}
    \begin{aligned}
        \hat{M}_r &= \ket{\psi_+}\!\bra{\psi_+} + \ket{\psi_-}\!\bra{\psi_-} \nonumber \\
        &+ e^{-dt/4\tau}\Bigg[\cosh\left(\frac{rdt}{2\tau}\right)\left(\ket{\phi_+}\!\bra{\phi_+} + \ket{\phi_-}\!\bra{\phi_-}\right) \\&+ \sinh\left(\frac{rdt}{2\tau}\right)\left(\ket{\phi_+}\!\bra{\phi_-} + \ket{\phi_-}\!\bra{\phi_+}\right)\Bigg],
    \end{aligned}
    \end{equation}
  that depends directly on the measured operator and its square,
  \begin{subequations}
  \begin{align}
  \label{eq:n-op}
      \hat{N} &= \frac{\hat{\sigma}_z^{(1)} + \hat{\sigma}_z^{(2)}}{2} = \ket{\phi_+}\!\bra{\phi_-} + \ket{\phi_-}\!\bra{\phi_+}, &\\
      \hat{N}^2 &= \ket{\phi_+}\!\bra{\phi_+} + \ket{\phi_-}\!\bra{\phi_-}, & \\
      \hat{N}^3 &= \hat{N}.
  \end{align}
  \end{subequations}
  Note that this state-dependent part can also be written compactly as 
  \begin{align}\label{eq:meas-op}
      \hat M_r = \exp[(dt/2\tau)(r \hat N -\hat N^2/2)], 
  \end{align}
  in agreement with the expected form of the Gaussian POVM in Eq.~\eqref{eq:povm-condition}. Unlike for single-qubit observable monitoring (e.g., \cite{pattiLinearFeedbackStabilization2017}), the $r$-independent quadratic term with $\hat{N}^2$ is nontrivial for the state update since it picks out the relevant projection for the subspace in which the measurement acts.
  
 	\subsection{Inefficient measurement}\label{sec:efficiency}
 	In a perfectly efficient measurement, the digitized measurement record is informationally complete, so after conditioning the state on the information in the collected record no information is lost and the state remains pure. However, a real measurement apparatus is not perfectly efficient due to signal loss during transit, as well as a degradation of the remaining information in the signal from additional noise due to thermal interactions, electronics noise, or the amplification chain. 
 	
 	The overall measurement efficiency can be characterized with a parameter $\eta \in [0,1]$, such that $\eta = 1$ corresponds to perfectly efficient measurement with no information loss, $\eta = 0$ corresponds to total information loss, and $0 < \eta < 1$ is a partially efficient measurement. Any loss of information increases the measurement time $\tau = 1/2\Gam \eta$ needed to integrate the signal to achieve a unit signal-to-noise ratio. The measurement operator in Eq.~\eqref{eq:meas-op} thus already accounts for this loss through the parameter $\tau$. 
  
    However, any lost information will be effectively averaged over in trajectory validation tomography, which will produce additional ensemble-average dephasing. In the limit of $\eta = 0$ with no collected signal, the best estimate one can claim is the ensemble-averaged evolution, which for Markovian noise should be in accordance with a Lindblad master equation. Thus, we include the additional ensemble-dephasing from information loss in each trajectory via the Lindblad map in Eq.~\eqref{eq:lindblad-map}, using $\hat J = \hat N/\sqrt{2}$ as the measured operator with a residual ensemble measurement-dephasing rate $(1-\eta)\Gam$,
 	\begin{subequations}
 		\begin{align}
 			\mathcal{L} [\rho] &= (1-\eta)(dt\,\Gam/2) \bigp{\hat N \rho \hat N} + \hat J_0 \rho \hat J_0^\dagger \\
 			\hat J_0 &=  \sqrt{\id - (1-\eta)(dt\,\Gam/2)\hat N^2}.
 		\end{align}
 		\label{eq:lindblad-meas}
 	\end{subequations}

 \section{Derivation of feedback protocols}
 The feedback task is to apply a control drive that maximizes the fidelity to the target state $\ket{\psi_+}$ while minimizing the fidelity with the invisible state $\ket{\psi_-}$. Such a control drive will be generated by the co- and counter-rotating drive Hamiltonians defined in the main text,
 	\begin{align}
 		\hat H_\Om(t) &= \Om\, \frac{\hat Y_{+}}{2},  &
 		\hat H_{\Delta}(t) &= \Del\, \frac{\hat Y_{-}}{2},
 	\label{eq:H-omega-delta}
 	\end{align}
 where $\hat Y_\pm \equiv \hat \sigma_y^1 \pm \hat \sigma_y^2$ and the frequencies $\Omega$ and $\Delta$ can generally depend upon the time $t$, the delayed and normalized measurement record $r(t-\tau_d)$, and/or the quantum state estimate $\rho_\text{est}(t)$.
 
  We use the standard (Uhlmann) definition for the state fidelity,
\begin{equation}
	\mathcal{F}(\sigma, \rho) \equiv \Bigp{\text{Tr} \sqrt{\sqrt \sigma ~\rho \sqrt \sigma}}^2,
\end{equation}
which for a pure target state $\sigma = \ket{\psi}\!\bra{\psi}$ and estimated state $\rho = \rho_\text{est}$ simplifies to,
\begin{equation}
	\mathcal{F}_\psi \equiv \mathcal{F}(\ket{\psi}\!\bra{\psi},~ \rho_\text{est}) = \braket{\psi |\rho_\text{est}|\psi},
\end{equation}
and can be interpreted as the population of $\rho_\text{est}$ in the target state $\ket{\psi}$.

 For constant $\Omega$ and $\Delta$, the unitary evolution generated by the control Hamiltonian $\hat H_\Om + \hat H_\Del$ in isolation over a time interval $dt$ is,
 \begin{equation}
	\begin{aligned}
		\label{eq:UGeneral}
		&\hat U_{\Om,\Del,dt} = ~ e^{-i(\Om\,\hat{Y}_+ + \Del\,\hat{Y}_-)dt/2} 
  \\& ~~~= e^{-i (\Om+\Del)\hat \sigma_y\,dt/2}\otimes e^{-i (\Om - \Delta)\hat \sigma_y\,dt/2} \\
		&~~~= ~ (|\psi_+\rangle\langle\psi_+| + |\phi_-\rangle\langle\phi_-|)\cos (\Om dt)
		\\ &\hspace{12mm}+ (|\psi_+\rangle\langle\phi_-| - |\phi_-\rangle\langle\psi_+|)\sin (\Om dt) \nonumber \\ &\hspace{12mm}+~ (|\phi_+\rangle\langle\phi_+| + |\psi_-\rangle\langle\psi_-|)\cos(\Delta dt)\\ &\hspace{12mm} +(|\phi_+\rangle\langle\psi_-| - |\psi_-\rangle\langle\phi_+|)\sin(\Delta dt), \nonumber  		
        \end{aligned}
        \end{equation}
where we note the following useful Bell state identities,
\begin{subequations}\label{eq:y-op}
\begin{align}
    \frac{\hat{Y}_+}{2i} ~~&= \ket{\psi_+}\!\bra{\phi_-} - \ket{\phi_-}\!\bra{\psi_+}, \\ 
    \left(\frac{\hat{Y}_+}{2}\right)^2 &= |\psi_+\rangle\langle\psi_+| + |\phi_-\rangle\langle\phi_-|, \\ \left(\frac{\hat{Y}_+}{2}\right)^3 &= \frac{\hat{Y}_+}{2}, \\ 
    \frac{\hat{Y}_-}{2i} ~~&= \ket{\phi_+}\!\bra{\psi_-} - \ket{\psi_-}\!\bra{\phi_+}, \\ \left(\frac{\hat{Y}_-}{2}\right)^2 &= |\phi_+\rangle\langle\phi_+| + |\psi_-\rangle\langle\psi_-|, \\ \left(\frac{\hat{Y}_-}{2}\right)^3 &= \frac{\hat{Y}_-}{2}, 
\end{align}
\end{subequations}
which imply the partition of unity,
\begin{equation}
\begin{aligned}
    \hat{I} &= \left(\frac{\hat{Y}_+}{2}\right)^2  +\left(\frac{\hat{Y}_-}{2}\right)^2 \\ &= \ket{\phi_+}\!\bra{\phi_+} + \ket{\phi_-}\!\bra{\phi_-} + \ket{\psi_+}\!\bra{\psi_+} + \ket{\psi_-}\!\bra{\psi_-}.
\end{aligned}
\end{equation}

For completeness, we also note that these operators are part of two natural sets of Pauli-like two-qubit operators. The other two in the $\{\ket{\psi_+},\ket{\phi_-}\}$ subspace are,
\begin{subequations}\label{eq:pauli-plus}
\begin{align}
    &\frac{\hat{X}_+}{2} = \frac{\hat{\sigma}_z^{(1)}\hat{\sigma}_x^{(2)} + \hat{\sigma}_x^{(1)}\hat{\sigma}_z^{(2)}}{2} = \ket{\psi_+}\!\bra{\phi_-} + \ket{\phi_-}\!\bra{\psi_+}, \\
    &\frac{\hat{Z}_+}{2} = \frac{\hat{\sigma}_z^{(1)}\hat{\sigma}_z^{(2)} - \hat{\sigma}_x^{(1)}\hat{\sigma}_x^{(2)}}{2} = \ket{\phi_-}\!\bra{\phi_-} - \ket{\psi_+}\!\bra{\psi_+}, \\
    &\left[\frac{\hat{X}_+}{2},\,\frac{\hat{Y}_+}{2}\right] = 2i\frac{\hat{Z}_+}{2} \quad \text{(and cyclic perm.)},
\end{align}
\end{subequations}
while the other two in the orthogonal $\{\ket{\psi_-},\ket{\phi_+}\}$ subspace are,
\begin{subequations}\label{eq:pauli-minus}
\begin{align}
    &\frac{\hat{X}_-}{2} = \frac{\hat{\sigma}_z^{(1)}\hat{\sigma}_x^{(2)} - \hat{\sigma}_x^{(1)}\hat{\sigma}_z^{(2)}}{2} = \ket{\psi_-}\!\bra{\phi_+} + \ket{\phi_+}\!\bra{\psi_-}, \\
    &\frac{\hat{Z}_-}{2} = -\frac{\hat{\sigma}_z^{(1)}\hat{\sigma}_z^{(2)} + \hat{\sigma}_x^{(1)}\hat{\sigma}_x^{(2)}}{2} = \ket{\psi_-}\!\bra{\psi_-} - \ket{\phi_+}\!\bra{\phi_+}, \\
    &\left[\frac{\hat{X}_-}{2},\,\frac{\hat{Y}_-}{2}\right] = 2i\frac{\hat{Z}_-}{2} \quad \text{(and cyclic perm.)}.
\end{align}
\end{subequations}

We now briefly derive two strategies for setting the control frequencies $\Omega$ and $\Delta$ so this control unitary maximizes the target fidelity.

\subsection{Optimal feedback without measurement or dephasing}
\label{subsec:optimal-deterministic}

We can derive a feedback protocol if we assume a known initial estimated state $\rho_\text{est}$ from which an optimal drive can be inferred. The updated state estimate after applying a control drive is $\rho'_\text{est} = \hat{U}_{\Om,\Del,\delta t}\,\rho_\text{est}\,\hat{U}^\dagger_{\Om,\Del,\delta t}$, which should be compared to the states $\ket{\psi_\pm}$.
For a pure target state $\sigma = \ket{\psi}\!\bra{\psi}$ and the post-control estimated state $\rho = \rho'_\text{est}$, the state fidelity has the form,
\begin{equation}
\begin{aligned}
	\mathcal{F}_\psi &\equiv \mathcal{F}(\ket{\psi}\!\bra{\psi},~ \rho'_\text{est}) \\&= \braket{\psi |\rho'_\text{est}|\psi} \\ &= \braket{\psi |\hat{U}_{\Om,\Del,\delta t}\,\rho_\text{est}\,\hat{U}^\dagger_{\Om,\Del,\delta t}|\psi}.
 \label{eq:fidelity-determinstic}
 \end{aligned}
\end{equation}
The state fidelities to $\ket{\psi_\pm}$ following a control drive $\hat U_{\Om, \Del, \delta t}$ are thus,
\begin{subequations}
	\begin{align}
		\mathcal{F}_{\psi+}(\Omega) &= \frac12 \Big(\rho_{\psi+}+\rho_{\phi-} + (\rho_{\psi+}-\rho_{\phi-})\cos(2\Om dt) \nonumber
  \\&+ 2\text{Re}(\rho_{\psi+,\phi-}) \sin(2\Om ~\delta t)\Big), \\
		\mathcal{F}_{\psi-}(\Delta) &=\frac12 \Big(\rho_{\psi-}+\rho_{\phi+} + (\rho_{\psi-}-\rho_{\phi+})\cos(2\Delta ~ \delta t) \nonumber
			 \\& - 2\text{Re}(\rho_{\phi+,\psi-}) \sin(2\Delta ~\delta t)\Big),
	\end{align}
\end{subequations}
where $\rho_{\psi\pm} = \braket{\psi_\pm | \rho_\text{est} |\psi_\pm}$, $\rho_{\phi\pm} = \braket{\phi_\pm | \rho_\text{est} |\phi_\pm}$, $\rho_{\psi+,\phi-} = \braket{\psi_+ | \rho_\text{est} |\phi_-}$, and $\rho_{\phi+,\psi-} = \braket{\phi_+ | \rho_\text{est} |\psi_-}$ are the relevant populations and coherences of the pre-control estimated state $\rho_\text{est}$. 
Because $\ket{\psi_+}$ is invariant under $\hat H_\Delta$, and $\ket{\psi_-}$ under $\hat H_\Omega$, their respective fidelities are only affected by one of the two drive parameters, $\Omega$ or $\Delta$. 

Extremizing these fidelities using the conditions $d\mathcal{F}_{\psi+}/d\Om = 0 = d\mathcal{F}_{\psi-}/d\Del$ then determines the optimal state-estimate-dependent constant drives to apply over a single control cycle $\delta t$,
\begin{subequations} \label{eq:OptRotation}
\begin{align}
		\tan\left[2\Om_\text{opt,0}(\rho_\text{est})\delta t\right] &= 2\Om_\text{opt,0}(\rho_\text{est})\delta t + \mathcal{O}(\delta t^3) \nonumber \\ &= \frac{2\text{Re}\rho_{\psi+,\phi-}}{\rho_{\psi+}-\rho_{\phi-}}, \\
		\tan\left[2\Delta_\text{opt}(\rho_\text{est})\delta t\right] &= 2\Del_\text{opt}(\rho_\text{est})\delta t + \mathcal{O}(\delta t^3) \nonumber \\ &= \frac{2\text{Re}\rho_{\phi+,\psi-}}{\rho_{\phi+}-\rho_{\psi-} }. \label{eq:DeltaOpt}
\end{align}
\end{subequations}
Because the right-hand-side of both expressions are $\mathcal{O}(1)$ in $\delta t$, which itself is an integer multiple of $dt$, the arctangent is not expanded in calculations of $\Omega_{\rm opt, 0}, \Delta_{\rm opt}$ used in simulations. Note that, unlike state-estimate-independent linear feedback schemes (e.g.,  Ref.~\cite{pattiLinearFeedbackStabilization2017}), the drive $\Omega\,\hat{Y}_+$ inherently disrupts the target state in our protocol, which forces the dependence on a state estimate to achieve optimal feedback control. This optimal feedback drive applies when the measurement backaction is ignored or negligible. In particular, the basic rate $\Omega_{\rm opt, 0}$ will acquire corrections when accounting for the measurement properly, as detailed in the next section.

For completeness, we also note that these optima can be written in terms of the observables in Eqs.~\eqref{eq:pauli-plus} and \eqref{eq:pauli-minus},
\begin{subequations} \label{eq:OptRotationPauli}
\begin{align}
		\Om_\text{opt,0}(\rho_\text{est})\delta t &= \frac12\tan^{-1} (C_+(\rho_{\rm est})),  \label{eq:OmegaOpt} \\
		\Del_\text{opt}(\rho_\text{est})\delta t &= \frac12 \tan^{-1}(C_-(\rho_{\rm est})),  \label{eq:DeltaOpt}
\end{align}
\end{subequations}
where
\begin{subequations}
    \begin{align}
        C_+(\rho) &= \frac{2\text{Re}\rho_{\psi+,\phi-}}{\rho_{\psi+}-\rho_{\phi-}} = -\frac{\langle\hat{X}_+\rangle}{\langle\hat{Z}_+\rangle}, \\
        C_-(\rho) &= \frac{2\text{Re}\rho_{\phi+,\psi-}}{\rho_{\phi+}-\rho_{\psi-} } = -\frac{\langle\hat{X}_-\rangle}{\langle\hat{Z}_-\rangle}.
    \end{align}
\end{subequations}

\subsection{Optimal feedback with measurement but no dephasing}
\label{subsec:optimal-measurement}

When concurrently monitoring an observable $\hat{N} = (\hat\sigma_z^{(1)} + \hat\sigma_z^{(2)})/2$, the feedback control will generally depend upon the delayed and normalized measurement record $r(t-\tau_d)$. As a minimal feedback protocol, the co-rotating drive should depend on the record and a state estimate $\rho_\text{est}$,
\begin{align}
    \Omega(t) &= \Omega[r(t-\tau_d),\,\rho_\text{est}(t)],
\end{align}
while the counter-rotating drive can be kept constant or independently optimized to the $\Delta_{\rm opt}(\rho_{\text est})$ in Eq.~\eqref{eq:DeltaOpt}. Combining the control drive in Eq.~\eqref{eq:UGeneral} with the measurement operator associated with $r(t-\tau_d)$ yields a composite update operator for the measurement interaction and evolution with delay time $\tau_d$ plus a following control time step $\delta t$ of $\hat{U}_{\Omega,\Delta,\delta t,r}\hat{U}_{\tau_d+\delta t}\hat{M}_{r}$. 

The time delay $\tau_d$ complicates the analysis, since the state generally evolves between when the measurement is performed and when the feedback control is applied. Distinguishing the control cycle timestep $\delta t$ from the evolution timestep $dt$ results in a similar complication, since even a $\tau_d = 0$ system will undergo evolution that goes uncorrected for an interval $\delta t$. However, as an initial approximation, we can take the zero-delay limit $\tau_d\to 0$ by ignoring the intermediate evolution $\hat U_{\tau_d+\delta t}$ before the control, and take the controller latency to be negligible by setting $\delta t = dt$. In this zero-delay case, the fidelity to the target state $\ket{\psi_+}$ after a measurement-and-control timestep $dt$ has the form,
\begin{align}\label{eq:fidelity-update}
    \mathcal{F}_{\psi+} &= \frac{\bra{\psi_+}\hat{U}_{\Om,\Del,dt,r}\hat{M}_r\,\rho_\text{est}\,\hat{M}_r^\dagger\hat{U}^\dagger_{\Om,\Del,dt,r}\ket{\psi_+}}{\text{Tr}[\hat{M}^\dagger_r\hat{M}_r\,\rho_\text{est}]}.
\end{align}
Treating the parameter $\Omega$ as tunable and using shorthand notation $\hat{M}=\hat{U}_{\Om,\Del,dt,r}\hat{M}_r$ for the composite operator, the extremization condition for this fidelity becomes,
\begin{align}\label{eq:fidelity-opt}
    \frac{\partial \mathcal{F}_{\psi+}}{\partial \Omega} &= 0 = 2\text{Re}\bra{\phi_-}\hat{M}\rho_\text{est}\hat{M}^\dagger\ket{\psi_+}.
\end{align}
Using the Bell state identities,
\begin{widetext}
\begin{subequations}
\begin{align}
    \hat{M}^\dagger\ket{\psi_+} &= \cos(\Omega dt)\ket{\psi_+} \sin(\Omega dt) e^{-dt/4\tau}\Bigg[\cosh\left(\frac{rdt}{2\tau}\right)\ket{\phi_-} + \sinh\left(\frac{rdt}{2\tau}\right)\ket{\phi_+}\Bigg], \\
    \hat{M}^\dagger\ket{\phi_-} &= -\sin(\Omega dt)\ket{\psi_+} + \cos(\Omega dt) e^{-dt/4\tau}\Bigg[\cosh\left(\frac{rdt}{2\tau}\right)\ket{\phi_-} + \sinh\left(\frac{rdt}{2\tau}\right)\ket{\phi_+}\Bigg],
\end{align}
\end{subequations}
the fidelity condition evaluates to a combination of components of $\rho_\text{est}$,
\begin{equation}
    \begin{aligned}
     2\text{Re}\bra{\phi_-}\hat{M}\rho_\text{est}\hat{M}^\dagger\ket{\psi_+} &= -\sin(2\Omega dt)\,\rho_{\psi+} + 2\cos(2\Omega dt)e^{-dt/4\tau}\Bigg[\cosh\left(\frac{rdt}{2\tau}\right)\text{Re}\rho_{\psi+,\phi-} + \sinh\left(\frac{rdt}{2\tau}\right)\text{Re}\rho_{\phi+,\psi+}\Bigg] \\
    &\qquad + \sin(2\Omega dt)e^{-dt/2\tau}\Bigg[\cosh^2\left(\frac{rdt}{2\tau}\right)\rho_{\phi-} + \sinh^2\left(\frac{rdt}{2\tau}\right)\rho_{\phi+} + \sinh\left(\frac{rdt}{\tau}\right)\text{Re}\rho_{\phi+,\phi-}\Bigg].
    \end{aligned}
\end{equation}
Setting this to zero, we can then solve directly for the optimal drive,
    \begin{align}
    \tan(2\Omega_{\rm opt} dt) &= \frac{2e^{-dt/4\tau}\left[\cosh\left(\frac{rdt}{2\tau}\right)\text{Re}\rho_{\psi+,\phi-} + \sinh\left(\frac{rdt}{2\tau}\right)\text{Re}\rho_{\phi+,\psi+}\right]}{\rho_{\psi+} - e^{-dt/2\tau}\left[\cosh^2\left(\frac{rdt}{2\tau}\right)\rho_{\phi-} + \sinh^2\left(\frac{rdt}{2\tau}\right)\rho_{\phi+} + \sinh\left(\frac{rdt}{\tau}\right)\text{Re}\rho_{\phi+,\phi-}\right]},
\\  & = \frac{2\text{Re}\rho_{\psi+,\phi-}}{\rho_{\psi+}-\rho_{\phi-}} + \frac{rdt}{2\tau}\left[\frac{2\text{Re}\rho_{\psi+,\phi+}}{\rho_{\psi+}-\rho_{\phi-}} + \frac{4\text{Re}\rho_{\psi+,\phi-}\text{Re}\rho_{\phi+,\phi-}}{(\rho_{\psi+}-\rho_{\phi-})^2}\right] \nn \\
  &\quad-\frac{dt}{4\tau}\left[\frac{\text{Re}\rho_{\psi+,\phi-}}{\rho_{\psi+}-\rho_{\phi-}} - \frac{4\text{Re}\rho_{\psi+,\phi+}\text{Re}\rho_{\phi+,\phi-} + 2\text{Re}\rho_{\psi+,\phi-}(\rho_{\phi+}-\rho_{\phi-})}{(\rho_{\psi+}-\rho_{\phi_-})^2} - \frac{2\text{Re}\rho_{\psi+,\phi-}(2\text{Re}\rho_{\phi+,\phi-})^2}{(\rho_{\psi+}-\rho_{\phi-})^3}\right] \nn \\ &\quad + \mathcal{O}(dt^{3/2}), \nonumber
\end{align}
where the second equality uses $(rdt)^2=\tau dt$.

Identifying the first term as the rescaled coherence $C_+(\rho)$ determining the measurement-free rate $\Omega_{\rm opt, 0}$ in Eq. \ref{eq:OmegaOpt} and recalling that $\tau=1/2\Gamma\eta$, this solution can be rewritten in the form,
\begin{equation}
    \begin{aligned}
        \label{eq:linearfeedback}
    \Omega_{\rm opt} dt &= \frac12 \tan^{-1}\Big(C_+(\rho_{\rm est})  + 2\eta\Gamma P(\rho_\text{est})\,rdt + \eta\Gamma Q(\rho_\text{est})\,\frac{dt}{2}\Big),
    \end{aligned}
\end{equation}
with the characteristic state-estimate-dependent factors,
\begin{subequations}
\begin{align}
    P(\rho_\text{est}) &= \frac{\text{Re}\rho_{\psi+,\phi+}}{\rho_{\psi+}-\rho_{\phi-}} + C_+(\rho_{\rm est})\frac{\text{Re}\rho_{\phi+,\phi-}}{\rho_{\psi+}-\rho_{\phi-}}, \label{eq:optimal-P}\\
    Q(\rho_\text{est}) &= -\frac{C_+(\rho_{\rm est}) }2 + \frac{4\text{Re}\rho_{\psi+,\phi+}\text{Re}\rho_{\phi+,\phi-} + 2\text{Re}\rho_{\psi+,\phi-}(\rho_{\phi+}-\rho_{\phi-})}{(\rho_{\psi+}-\rho_{\phi_-})^2}  + \frac{2\text{Re}\rho_{\psi+,\phi-}(2\text{Re}\rho_{\phi+,\phi-})^2}{(\rho_{\psi+}-\rho_{\phi-})^3}.
\end{align}
\end{subequations}
\end{widetext}
 The dominant terms are the measurement-free rate $C_+(\rho)$, of order 1, followed by the white noise term of order $dt^{1/2}$ in $rdt = 2\text{Re}\rho_{\phi+,\phi-}dt + dW/\sqrt{2\Gamma\eta}$.  Expanding the arctangent about the $\mathcal{O}(1)$ term $C_+(\rho_{\rm est})$ and keeping terms of order $dt^{3/2}$ gives
 \begin{subequations}
     \begin{align}
         \Omega_{\rm opt} dt &\approx \frac12 \tan^{-1} (C_+(\rho_{\rm est})) 
         + \frac{\eta\Gamma P(\rho_\text{est})\,rdt}{1 + C_+(\rho_{\rm est})^2} \\
         &=  \Omega_{\rm opt,0}(\rho_{\rm est}) + \eta \Gamma \Bigp{\frac{P(\rho_\text{est})\,}{1 + C_+(\rho_{\rm est})^2}} rdt.
     \end{align}
 \end{subequations}
 Thus, incorporating the most recent measurement information $rdt$ leads to a \textit{linear} correction, multiplied by a state-dependent factor. In the case where the coherence $\rho_{\psi+,\phi-}$ is being consistently and successfully rotated into the $\rho_{\psi_+}$ population, $C_+(\rho_{\rm est}) \approx 0$ and the feedback is linear up to the state-estimation for $P(\rho_{\rm est})$:
 \begin{equation}
      \Omega_{\rm opt} dt \approx \eta\Gamma P(\rho_\text{est})\,rdt.
      \label{eq:only-linear}
 \end{equation}

Note that $P(\rho_{\rm est})$ can also be written in terms of the operators in Eqs.~\eqref{eq:pauli-plus}, \eqref{eq:pauli-minus}, and \eqref{eq:n-op} as,
\begin{equation}
     P(\rho_\text{est}) = -\frac{\langle\hat{X}_+\hat{N}+\hat{N}\hat{X}_+\rangle}{4\langle\hat{Z}_+\rangle} + \frac{2\langle\hat{X}_+\rangle\langle\hat{N}\rangle}{\langle\hat{Z}_+\rangle^2}. \label{eq:optimal-P-2}
\end{equation}
This means that
the two dominant terms depend mainly on the pre-measurement coherence $\braket{\hat X_+}$ and the measurement-disturbed coherence $\langle\hat{X}_+\hat{N}+\hat{N}\hat{X}_+\rangle/2$, both weighted by population bias $\langle\hat{Z}_+\rangle$ in the subspace. This further motivates thinking of Eq. \ref{eq:only-linear} as the optimal drive when the feedback is on track: given an initial condition $\rho_0$ with $\braket{\hat X_+} = 0$, the drive following a measurement will just be $P(\rho_0) = -\bigp{\langle\hat{X}_+\hat{N}+\hat{N}\hat{X}_+\rangle} /{4\langle\hat{Z}_+\rangle}$. Since this drive is optimal, it should rotate all measurement-disturbed coherence into $\ket{\psi_+}$, thus restoring $\braket{\hat X_+} = 0$ until the next measurement. However, the introduction of nonidealities (time-delay and fluctuator / white noise) may cause the trajectory to deviate from this optimal feedback, making the $C_+(\rho_{\rm est})$ terms critical to the feedback success. This is explored further in Sec.~ \ref{subsec:correspondence}.

%
%
%
%
 
\subsection{Optimal feedback with measurement and dephasing}

The preceding section neglected the average dephasing occuring during the timestep $dt$ prior to the control being applied. However, this dephasing can be readily included in principle. Since the specifics of the dephasing are unknown to the feedback controller, the fidelity optimization condition of Eq.~\eqref{eq:fidelity-opt} should be modified to,
\begin{align}\label{eq:fidelity-opt-dephase}
    \frac{\partial \mathcal{F}_{\psi+}}{\partial \Omega} &= 0 = 2\text{Re}\bra{\phi_-}\mathcal{L}_{dt}(\hat{M}\rho_\text{est}\hat{M}^\dagger)\ket{\psi_+},
\end{align}
with the average dephasing modeled by a Lindblad state update,
\begin{subequations}
\begin{align}
    \mathcal{L}_{dt}(\rho) &= \frac{dt\Gamma_2}{2}\left(\hat{\sigma}_z^{(1)}\rho\hat{\sigma}_z^{(1)} + \hat{\sigma}_z^{(2)}\rho\hat{\sigma}_z^{(2)}\right) \nn \\ &\quad + \frac{(1 - \eta)dt\Gamma}{2}\hat{N}\rho\hat{N} + \hat{J}_0\rho\hat{J}_0^\dagger, \\
    \hat{J}_0^2 &= \id - \frac{dt\Gamma_2}{2}(2) - \frac{(1-\eta)dt\Gamma}{2}\hat{N}^2,
\end{align}
\end{subequations}
that includes both the effective single-qubit dephasing and the residual measurement dephasing due to imperfect efficiency $\eta<1$. However, including this dephasing over the single timestep $dt$ will only add corrections of order $dt$ to the deterministic feedback function $Q(\rho_\text{est})$ in Eq.~\eqref{eq:linearfeedback}, which is already subdominant, so the dephasing corrections can also be safely neglected.

 \subsection{Linear feedback stochastic Schr\"odinger equation}
 \label{app:linearFeedback}
 Assuming ideally efficient linear feedback in the form of Eq.~\eqref{eq:linearfeedback}, we can also derive an effective stochastic Schr\"odinger equation that describes the feedback dynamics. For simplicity, consider feedback drives of the minimal form,
 \begin{align}
     \Omega(t) &= \Gam P(\rho_\text{est}) r(t), & \Delta &= \Gam c, 
 \end{align}
 with a common scaling constant $\Gam$ set to match the ensemble-average measurement-dephasing rate. For the co-rotating drive, the tunable state-estimate-dependent function $P(\rho_\text{est})$ scales the measurement record, again assuming zero time delay for convenience, while the strength of the counter-rotating drive can be relatively tuned with the dimensionless constant $c$. 

Following the form of the update in Eq.~\eqref{eq:fidelity-update} the renormalized state after both measurement and the unitary feedback drive is,
\begin{equation}
\ket{\psi(t+dt)} = \frac{ \hat U_{\Omega,\Delta,dt,r}\hat M_{r}}{\sqrt{\braket{\psi | \hat M_{r}^\dagger  \hat M_{r}|\psi}}} \ket{\psi}.
    \label{eq:kraus-map}
\end{equation} 
In this zero-delay case, using the white noise approximation for the readout in Eq.~\eqref{eq:wiener}, the form of the measurement operator in Eq.~\eqref{eq:meas-op}, and the form of the control unitary in Eq.~\eqref{eq:UGeneral} yields the following composite update map to order $dt$, recalling that $(r\,dt)^2 = \tau\,dt +\mathcal{O}(dt^{3/2})$,
\begin{equation}
    \begin{aligned}
    &\frac{\hat{U}_{\Omega,\Delta,dt,r}\hat{M}_r}{\sqrt{\langle\hat{M}_r^\dagger\hat{M}_r\rangle}} \approx  \\
    &\quad \approx \hat{I} + (r\,dt)\left[\frac{\hat{N}-\langle\hat{N}\rangle}{2\tau} + \Gamma P\,\frac{\hat{Y}_+}{2i}\right] \\ & \qquad + dt\left[\Gamma c\,\frac{\hat{Y}_-}{2i} + \frac{\Gamma P}{2}\frac{\hat{Y}_+}{2i}(\hat{N}-\langle\hat{N}\rangle)\right] \\
    & \qquad - \frac{dt}{2}\left[\frac{\hat{N}^2 +2\hat{N}\langle\hat{N}\rangle - 3\langle\hat{N}\rangle^2}{4\tau} - \tau(\Gamma P)^2\left(\frac{\hat{Y}_+}{2i}\right)^2\right].
    \end{aligned}
\end{equation}

After substituting the white noise relation $r\,dt = \langle\hat{N}\rangle dt + \sqrt{\tau}\,dW$ and writing the measurement time $\tau = 1/2\Gamma$ in terms of the dephasing rate $\Gamma$ assuming perfect efficiency $\eta=1$, this yields a stochastic Schr\"odinger equation that includes (instantaneous and perfectly efficient) feedback,
\begin{equation}
    \begin{aligned}
            d\ket{\psi} &= \sqrt{\frac{\Gamma}{2}}\,dW\left[(\hat{N}-\langle\hat{N}\rangle) - i \frac{P}{2}\,\hat{Y}_+\right]\ket{\psi}  \\ 
            &\quad - i\frac{\Gamma}{2}\,dt\left[ c\,\hat{Y}_- + \frac{P}{2}\hat{Y}_+(\hat{N}+\langle\hat{N}\rangle)\right]\ket{\psi} \\
    &\quad - \frac{\Gamma}{2}\,dt\left[\frac{\left(\hat{N} - \langle\hat{N}\rangle\right)^2}{2} + \frac{P^2}{8}\hat{Y}_+^2\right]\ket{\psi}.
    \end{aligned}
\end{equation}

\subsection{Correspondence of deterministic and record-dependent feedback}
\label{subsec:correspondence}

The optimal protocols derived in Secs. \ref{subsec:optimal-deterministic} and \ref{subsec:optimal-measurement} are equivalent under a special choice of state estimate. In particular, Eq. \ref{eq:fidelity-update} can be rearranged as
\begin{subequations}
    \begin{align}
    \mathcal{F}_{\psi+} &= \bra{\psi_+}\hat{U}_{\Om,\Del,dt,r}\Biggp{\frac{\hat{M}_r\,\rho_\text{est}\,\hat{M}_r^\dagger}{\text{Tr}[\hat{M}^\dagger_r\hat{M}_r\,\rho_\text{est}]}}\hat{U}^\dagger_{\Om,\Del,dt,r} \ket{\psi_+} \\
    &= \bra{\psi_+}\hat{U}_{\Om,\Del,dt,r}\rho_{\rm est}'\hat{U}^\dagger_{\Om,\Del,dt,r} \ket{\psi_+},
\end{align}
\end{subequations}
which is equivalent to using the Bayesian-updated state $\rho_{\rm est}'$ for the deterministic feedback in Eq. \ref{eq:fidelity-determinstic}. Thus, the optimal feedback that includes the linear feedback term (Eq. \ref{eq:linearfeedback}) is implicitly realized by applying deterministic feedback (Eq. \ref{eq:OmegaOpt}) on the Bayesian-updated state.

This has important practical consequences. The fact that the measurement-dependent feedback requires state-tracking mitigates its potential advantage over the deterministic feedback. In a state-independent linear feedback protocol such as \cite{vijayStabilizingRabiOscillations2012}, feedback can be implemented in hardware, eliminating computational overhead. However, since state-estimation is required with both record-dependent and record-independent expressions we derived, it is optimal to compute the computationally simpler deterministic expression Eq. \ref{eq:fidelity-determinstic} after performing the Bayesian update. In this case, the record-dependence enters the feedback implicitly through the Bayesian update.

\begin{figure}
    \centering
    \includegraphics[width=\linewidth]{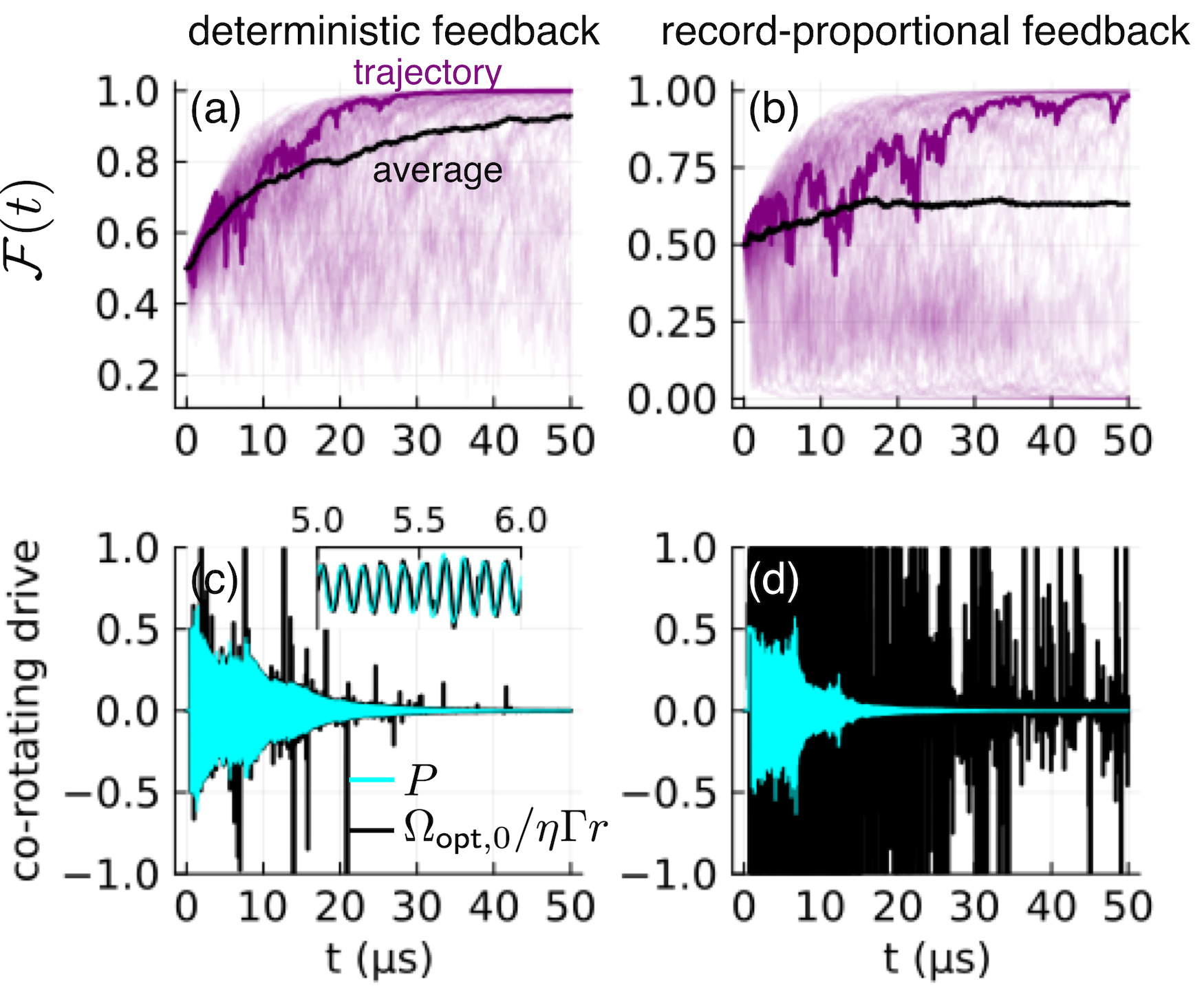}
    \caption{Fidelity $\mathcal{F}(t)$ to target $\ket{\psi_+}$, comparing deterministic feedback $\Omega_{\rm opt,0}(\rho)$ (left) to record-proportional feedback $\Omega_{\rm opt,1}(\rho) = \eta\Gamma\,P(\rho)\,r$ (right), with $1/f$-noise $\Gamma_2 = 1/52\mu$s, $\Delta_{\rm d.d.}/2\pi = 10$~MHz, $\eta = 1/2$, $\tau_d = 500$~ns, and $\Gamma = 1/\mu$s.
  $N = 100$ trajectories 
  \textbf{(a,~b)} Overlaid ensemble of $100$ trajectories (translucent purple) highlighting one trajectory (dark purple) and the ensemble-average (black). Deterministic feedback in \textbf{(a)} stabilizes the ensemble-mean over time, while proportional feedback in \textbf{(b)} with a delayed record $r(t-\tau_d)$ does not reliably stabilize all trajectories. 
  \textbf{(c,~d)} Comparison of the deterministic drive $\Omega_{\rm opt,0}[\rho(t)]/\eta\Gamma r(t-\tau_d)$ scaled by the delayed record (black) and the envelope $P[\rho(t-\tau_d)]$ used in proportional feedback with no forward-estimation (aqua). For deterministic feedback in \textbf{(c)}, the drive $\Omega_\text{opt,0}$ is computed and used to get the highlighted trajectory in (a) first, then $P$ is computed for that observed trajectory and shows close agreement to the deterministic envelope (inset). For proportional feedback in \textbf{(d)}, the $P$ is computed and used to get the highlighted and stabilized trajectory in (b) first, then $\Omega_\text{opt,0}$ is computed but shows poor agreement.}
    \label{fig:proportional}
\end{figure}

For completeness, we compare simulations of two approaches in Fig. \ref{fig:proportional}. The left column depicts deterministic feedback, which uses only the $\Omega_\text{opt,0}$ term from Eq. \ref{eq:OmegaOpt}. The right column depicts record-proportional feedback, which uses only the first term in Eq. \ref{eq:optimal-P}, i.e.
\begin{equation}
    \begin{aligned}
         P(\rho_\text{est}) &= \frac{\text{Re}\rho_{\psi+,\phi+}}{\rho_{\psi+}-\rho_{\phi-}} + \frac{2\text{Re}\rho_{\psi+,\phi-}\text{Re}\rho_{\phi+,\phi-}}{(\rho_{\psi+}-\rho_{\phi-})^2} \\
         &\approx  \frac{\text{Re}\rho_{\psi+,\phi+}}{\rho_{\psi+}-\rho_{\phi-}}.
    \end{aligned}
\end{equation}
This is chosen on the basis that the coherence $\text{Re}\rho_{\psi_+,\phi-}$ and the measured $\langle\hat{N}\rangle$ are proportional to $\sqrt{\rho_{\phi-}}$ so become small near $\ket{\psi_+}$. Including $r$, the noise will dominate over deterministic terms in Eq. \ref{eq:linearfeedback}, so $\Omega_{\rm opt}dt \approx \eta\Gamma\,P[\rho]\,rdt$ so long as the feedback remains close to the optimal trajectory.


The deterministic feedback in \ref{fig:proportional}a stabilizes more reliably than the proportional feedback in \ref{fig:proportional}b. This makes sense given that the deterministic feedback includes the zero-order term while also including record information implicitly, whereas the proportional feedback excludes the zero-order term. Interestingly, for the highlighted trajectory stabilized with deterministic feedback, the forward-estimated drive $\Omega_{\rm opt,0}[\rho_\text{est}(t)]/\eta\Gamma r(t-\tau_d)$ scaled by the delayed record in \ref{fig:proportional}c (black) is in close agreement with the $P[\rho_\text{est}(t-\tau_d)]$ that would have been calculated for that trajectory \emph{without forward-estimation} (aqua). This supports the intuition that the deterministic feedback implicitly includes the record information, and also confirms that the proportional term is the dominant effect most of the time. However, for a trajectory stabilized with proportional feedback only, the used $P$ in \ref{fig:proportional}d (aqua) shows poor agreement with the drive $\Omega_{\rm opt,0}/\eta\Gamma r$ (black) that would have been calculated. This suggests that the infrequent, large discrepancies between $\Omega_\text{opt,0}$ and $P$ in Fig.~\ref{fig:proportional}c represent nonlinear corrections that are critical to the feedback success. Without these large corrections in the proportional feedback case, the deterministic protocol always predicts a large correction needed to return to the more stable regime where proportional feedback becomes reliable.

\section{Improving feedback control with forward state-estimation}
\label{app:forward-estimation}
The feedback control derived in the previous section ignored the time delay $\tau_d$ between when the measurement record is collected and the control is applied. However, in practice this time lag is necessary for information to travel between the system and measurement apparatus, so concurrent system evolution will take place during this delay that should not be neglected in the state estimate $\rho_{\rm est}$ being used to predict the feedback control that should be applied. Thus, the applied feedback control should anticipate what the \emph{future state} will be, accounting for the evolution between when the measurement interaction occurred in the past to produce the collected record and when the feedback signal reaches the system.

While unitary control drives are known and may be readily accounted in such a \emph{forward estimation} of the future state, other types of evolution like measurement and fluctuations are not known during the delay. As such, the best estimate one can make for the evolution during the delay is the \emph{mean evolution} that averages over the unknowns and uses the information about what drives have already been output by the controller but have not yet arrived at the system.
  

The feedback loop time delay $\tau_d$ between when the measurement microwave field interacts with the system and when the corresponding feedback control field reaches the system can be broken into three conceptual parts,
\begin{align}
    \tau_d &= \tau_{s\to c} + \tau_{c} + \tau_{c\to s},
\end{align}
where $\tau_{s\to c}$ is the time of flight for the readout to make it through the measurement chain to the controller, $\tau_{c}$ is the time it takes for the controller to compute the feedback to apply, and $\tau_{c\to s}$ is the time of flight for the control signal to reach the system. Thus, the controller should understand the readout received at time $t$ as $r(t-\tau_{s\to c})$, meaning it reflects information in the system state at a time $\tau_{s\to c}$ in the past, but it should issue a control signal that is appropriate for the future system state $\rho(t - \tau_{s\to c} + \tau_d) = \rho(t + \tau_c + \tau_{c\to s})$. However, this distinction is unimportant for the controller, which can operate with a shifted internal clock time $t\mapsto t+\tau_{s\to c}$ so that its internal $t$ is matched to the time of the measurement interaction with the past system state, in which case it can assume that it receives a signal $r(t)$ and must output a feedback signal for the future state $\rho(t + \tau_d)$ a full delay time $\tau_d$ in the future.

Other than this full delay time $\tau_d$, the only important timescale for the feedback is the processing time of the controller $\tau_c$. This time sets the minimum duration required to change the output feedback control signal. Whereas the readout of the system is modeled as a discrete process with timestep $dt$ according to Eq. \ref{eq:discrete-r}, the readout processed by the controller is buffered and averaged in larger bins of duration $\tau_c$ to produce the values $\tilde r(t)$. In this way, no signal is lost during the processing time, but the controller is nevertheless restricted to outputting a signal that is time-stepped into bins no finer than its processing time $\tau_c$. For the purposes of this work, we assume that this processing time is what sets the overall control timestep $\delta t = \tau_c$ in the analysis.

With the shifted clock, the feedback controller should track an internal state estimate $\rho_\text{est}(t)$ corresponding to the actual system state $\rho_\text{sys}(t)$ that produced the received readout $\tilde r(t)$. It should then assume that the feedback signal will be applied to the appropriate \emph{forward estimate} of the system state,
\begin{subequations}
\label{eq:forwardestimate}
    \begin{align}
   \rho_\text{est}(t+\tau_d) &= \prod_{k=0}^{\lfloor\tau_d/\delta t\rfloor}\mathcal{E}_{\delta t}(t+ k \delta t)[\rho_\text{est}(t)], \\
   \mathcal{E}_{\delta t}(t_k) &= \mathcal{L}_{\delta t}\circ\mathcal{U}_{\delta t}(t_k),
    \end{align}
\end{subequations}
that has been evolved forward by $\lfloor\tau_d/\delta t\rfloor$ timesteps $\delta t$, including both known unitary dynamics $\mathcal{U}_{\delta t}(t_k)[\cdot] = \hat{U}_{\delta t}(t_k)(\cdot)\hat{U}^\dagger_{\delta t}(t_k)$ for the drives that were already sent by the controller during the preceding $\tau_d$ (which have not yet arrived at the system) and the average deocherence $\mathcal{L}_{\delta t}$ that is anticipated during each timestep $\delta t$ over the delay time $\tau_d$. 

In practice, the average decoherence is relatively small if the delay $\tau_d$ is sufficiently short, so can be neglected as an initial approximation. In this case, the forward estimation is entirely determined by the sequence of control unitaries for the drives that the controller has already sent. Since the unitary generators $\hat{Y}_\pm$ commute, the controller only needs to save a running buffer sum of the emitted drives, $(\Omega(t),\Omega(t-\delta t),\ldots,\Omega(t-\tau_d))$ and $(\Delta(t),\Delta(t-\delta t),\ldots,\Delta(t-\tau_d))$, from which it can keep running sums $\Omega_{\tau_d}$ and $\Delta_{\tau_d}$ over each buffer by subtracting the oldest and adding the newest when the buffer is updated at each time step. The forward-estimation unitary operator used for simulations in the main text is then simply,
\begin{align}
    \hat{U}_{\tau_d} &= \exp\left(-i(\Omega_{\tau_d}\hat{Y}_+ + \Delta_{\tau_d}\hat{Y}_-)\delta t/2\right), 
\end{align}
with the corresponding closed form in Eq.~\eqref{eq:UGeneral}. This sort of buffering and calculation is straightforward and fast on an FPGA controller.

 	\bibliography{main,appendix}

\end{document}